\documentclass{article}
\usepackage{graphicx}
\usepackage{amsmath,amssymb,amsthm,mathrsfs}
\usepackage{geometry}
\usepackage{ragged2e}
\newcommand*\pFq[2]{{}_{#1}F_{#2}}

\title{Rigorous results for mean first-passage time of harmonically trapped
particle}
\author{Przemys\l{}aw Che\l{}miniak}
\date{}

\begin{document}
\maketitle

\vspace{-0.8cm}
\begin{center}
{\it Institute of Spintronics and Quantum Information, Faculty of Physics, Adam
Mickiewicz University, Uniwersytetu Pozna\'nskiego 2, 61-614 Pozna\'n, Poland}
\end{center}

\vspace{-2mm}
\vspace{1mm}
\begin{abstract}
\noindent The Ornstein-Uhlenbeck process of diffusion in the harmonic potential
is re-examined in the context of the first-passage time problem. We investigate
this problem to the extent that it has not yet been fully resolved and
demonstrate exact novel results. They mainly concern the mean first-passage time
for a particle diffusing downward and upward in the harmonic potential. We
verify the main results of this paper by using a number of analytical techniques.
\end{abstract}

\section*{1.~Introduction}

The first-passage time distribution belongs to important physical quantities
describing the properties of diffusive motion in terms of spatiotemporal
relationships~\cite{Redn2001,Osha2014}. This statement refers to freely diffusing
objects as well as those executing stochastic motion in confining potentials
~\cite{Greb2015}. From a viewpoint of physical sciences, the harmonic potential
plays a relevant role. Theoreticians implement it very eagerly in simple models, 
which they treat as the first approximation of much more advanced theories.
In addition to that, these models are strictly solvable on the whole. For
experimentalists, the harmonic potential is relatively easy to set up in their
laboratories applying high-tech apparatus, even though optical tweezers
~\cite{Bust2021}. In addition, most trapping potentials can be successfully
approximated to the harmonic potential in the vicinity of its bottom. 

In this paper, we merge diffusive dynamics of a single (colloidal) particle with
the harmonic potential, in which this process takes place. Recall that such a model
was primarily studied by L.~S.~Ornstein and G.~E.~Uhlenbeck, and until today is
known by their names~\cite{Orns1930}. Here, however, we explore it anew paying
special attention to the first-passage time problem. More precisely, we consider
the average time required for a particle to reach a predetermined target, which
in formal terms is the first moment of the first-passage time distribution~\cite{
Gard2004}. For this reason, it is called the mean first-passage time, hitting time,
crossing time, or exit time, depending on the context. In the case of a freely
diffusing particle, the mean first-passage turns out to be infinite even though the
particle is sure to ultimately hit the target. In principle, we have no reason to
question such a result, due to the fact that the first-passage time distribution
is, by definition, normalized to unity. By contrast, when a particle diffuses in
bounded domains, its mean first-passage time from some initial position to the
target point is finite. The same regularity refers to diffusive motion confined
by external potentials, an example of which is the harmonic potential as in the
case of the Ornstein-Uhlenbeck process. While it is widely known that the mean
first-passage time should also be finite in the harmonic potential, we still do
not have, to our best knowledge, the exact analytical formula for this quantity,
albeit some partial solutions have been already obtained~\cite{Gupt2021,
Sant2021}. Therefore, the central objective of the present paper is to find the
rigorous solution for the mean first-passage time, both downward and upward of
the harmonic potential.

Let us recall that the overdamped motion of a single particle constrained by
one-dimensional potential $V(x)$ is described in terms of the Langevin equation
\begin{equation}
\frac{\text{d}x(t)}{\text{d}t}=-\mu\frac{\text{d}V(x)}{\text{d}x}+\xi(t),
\label{eq_1}
\end{equation}
in which the co-ordinate $x(t)$ indicates a position of the particle at time $t$
and $\mu$ determines its mobility~\cite{Mahn2009}. Thermal fluctuations
(Gaussian white noise) $\xi(t)$ with the average $\left<\xi(t)\right>\!=\! 0$
obeys the correlation relation $\left<\xi(t)\xi(t')\right> \!=\!2D\delta(t-t')$,
where $D$ is the diffusion coefficient. In equilibrium, the diffusion coefficient
$D$ and the mobility $\mu$ are related by the Einstein fluctuation-dissipation
theorem, $D\!=\!\mu k_{\text{B}}T$, with $k_{\text{B}}$ and $T$ standing for,
respectively, the Boltzmann constant and absolute temperature.

\begin{figure}[t]
\centering
\includegraphics[scale=0.25]{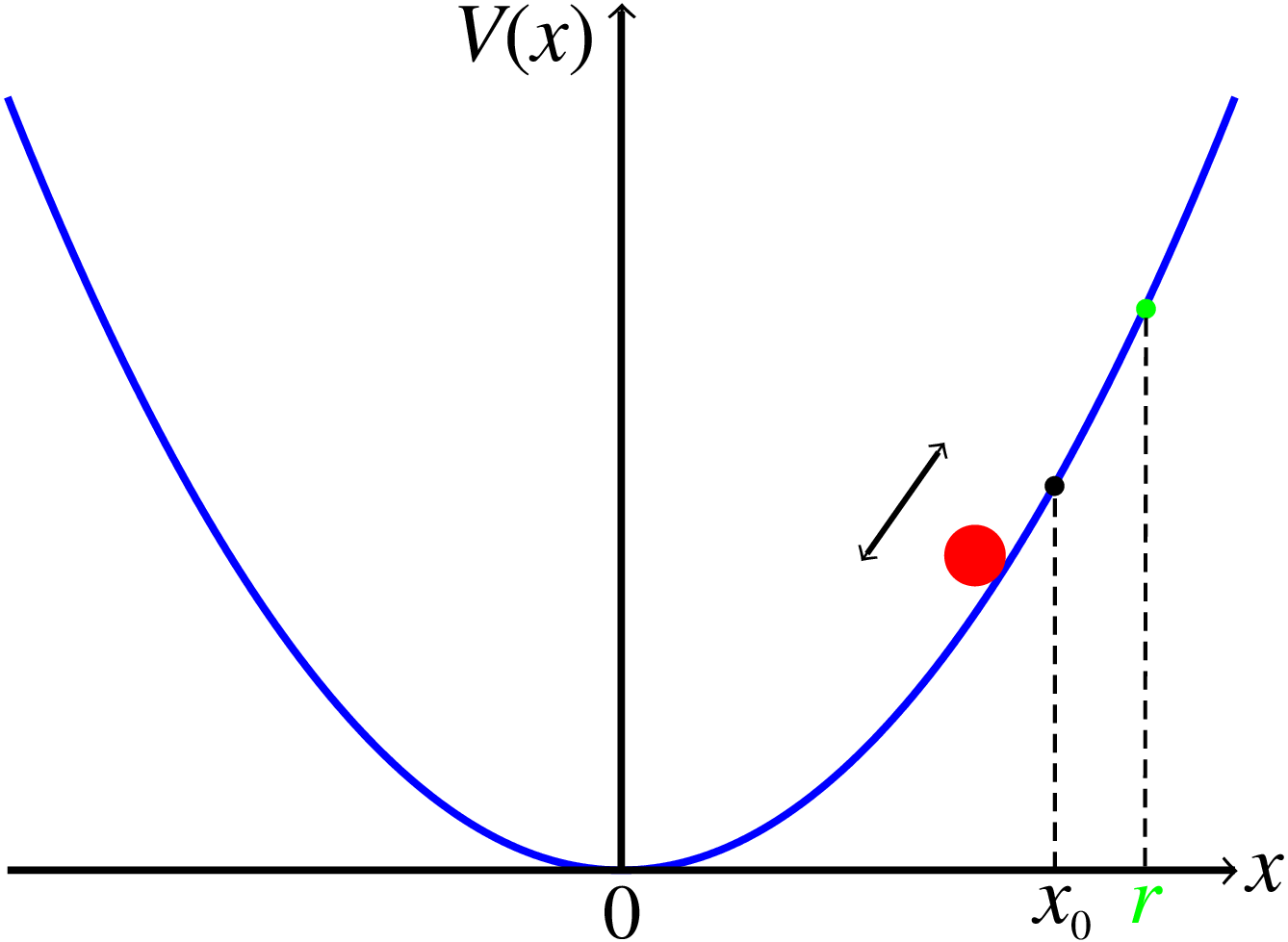}
\caption{Diffusion of a particle (red circle) in the harmonic potential. The 
question concerns the mean first-passage time of the particle from its initial
position $x_{0}\!>\!0$ downward to the target in the minimum $x\!=\!0$ of the
potential and vice versa from the initial position at $x\!=\!0$ upward to the
target localized at $x_{0}\!>\!0$. The green point at the position $x\!=\!r$
symbolizes the reflecting barrier.}
\label{fig1}
\end{figure}

The stochastic differential equation (\ref{eq_1}) is equivalent to the
Smoluchowski (Fokker-Planck) partial differential equation
\begin{equation}
\frac{\partial}{\partial t}p(x,t\!\mid\!x_{0})=\frac{\partial}{\partial x}
\!\left(\mu\frac{\mathrm{d} V(x)}{\mathrm{d} x}p(x,t\!\mid\!
x_{0})+D\frac{\partial}{\partial x}p(x,t\!\mid\!x_{0})\right),
\label{eq_2}
\end{equation}
which describes deterministic evolution of the probability distribution
$p(x,t\!\mid\!x_{0})$, also called the propagator, to observe the particle at
position $x$ at time $t$, given that initially it was localized at position
$x_{0}$~\cite{Risk1989}. In the particular case of the harmonic potential
$V(x)\!=\!\frac{1}{2}k x^{2}$ with the strength or stiffness parameter $k$,
Eq.~(\ref{eq_2}) takes the form
\begin{equation}
\frac{\partial}{\partial t}p(x,t\!\mid\!x_{0})=\alpha\frac{\partial}{\partial x}
\!\left[x\,p(x,t\!\mid\!x_{0})\right]+D\frac{\partial^{2}}{\partial x^{2}}
p(x,t\!\mid\!x_{0}),
\label{eq_3}
\end{equation}
into which we have introduced the new parameter $\alpha\!=\!k\mu$. The above
equation models in formal terms the Ornstein-Uhlenbeck process which corresponds
to the diffusion of the particle trapped by the harmonic potential. The exact
solution of Eq.~(\ref{eq_3}) augmented by the initial condition
$p(x,0\!\mid\!x_{0})\!=\!\delta(x-x_{0})$ is very well known and reads
\begin{equation}
p(x,t\!\mid\!x_{0})=\sqrt{\frac{\alpha}{2\pi D\!\left(1-\mathrm{e}^{-2\alpha t}
\right)}}\exp\!\left[-\frac{\alpha\,(x-x_{0}\,\mathrm{e}^{-\alpha t)^{2}}}{2D\!
\left(1-\mathrm{e}^{-2\alpha t}\right)}\right].
\label{eq_4}
\end{equation}
This probability distribution usually constitutes the input data for further
calculations concerning the mean first-passage time for diffusion downward of the
harmonic potential from the initial position $x_{0}\!>\!0$ to its minimum at
$x\!=\!0$. Fig.~(\ref{fig1}) more precisely visualizes the problem we study in
this paper. The inverse problem of finding the mean first-passage time upward of
the harmonic potential requires another approach, which will also be discussed
in the subsequent sections.

The structure of the paper is as follows. In Sec.~2 the first-passage time
statistics is briefly outlined and the analytical techniques we use in this
paper are indicated. Sec.~3 presents derivations and exact results for the mean
first-passage time downhill of the harmonic potential. In turn, the exact formula
for the mean first-passage time uphill of this potential is derived in Sec.~4.
The paper is summarized in Sec.~5.

\section*{2.~First-passage time statistics}

In this section we restrict our considerations to one spatial dimension, although
a generalization to higher dimensions is straightforward. In the subsequent
section, the first-passage time problem for diffusion in the harmonic potential
will be considered only in the one-dimensional case.

The mean first-passage time $\mathcal{T}(x_{0}\!\to\!x_{\mathrm{f}})$ from the
initial position $x_{0}$ to some prescribed target $x_{\mathrm{f}}$ (a final
point) is defined as the first moment of the first-passage time distribution
$F(t\!\mid\!x_{0})$. In formal terms
\begin{equation}
\mathcal{T}(x_{0}\!\to\!x_{\mathrm{f}})=\int_{0}^{\infty}t\,F(t\!\mid\!x_{0})\,
\mathrm{d}t,
\label{eq_5}
\end{equation}
where the function
\begin{equation}
F(t\!\mid\!x_{0})=-\frac{\mathrm{d}}{\mathrm{d}t}Q(t\!\mid\!x_{0})
\label{eq_6}
\end{equation}
relates to the time derivative of the cumulative survival probability $1\!-\!
Q(t\!\mid\!x_{0})$. Here, $Q(t\!\mid\!x_{0})\!=\!\int_{0}^{L}p(x,t\!\mid\!x_{0})
\mathrm{d}x$ is the survival probability of the particle that diffuses in the
interval of the length $L$. On the other hand, if we postulate $L\!=\!\infty$
then diffusion proceeds in the semi-infinite interval, which is also permissible
and may be in the realm of our interests. In any case, this quantity estimates
a chance that a diffusing particle survives until time $t$ (or not as predicted
by the cumulative survival probability) before reaching either end $x_{\mathrm{f}}
\!=\!0$ or $L$ of the interval $(0,L)$ for the first time. For diffusion along
the semi-infinite interval $(0,\infty)$, the target $x_{\mathrm{f}}$ is usually
established at its origin $x\!=\!0$. The survival probability manifests three
essential properties. At the initial position $x_{0}$ coinciding with the 
endpoints of the interval $(0,L)$, i.e. when $x_{0}\!=\!0$ or $x_{0}\!=\!L$,
$Q(t\!\mid\!x_{0})\!=\!0$ for any time $t\!>\!0$, which means that the particle
had no chance of survival being at the target from the very beginning or remains
there forever if both ends of the interval act as absorbing traps. At time
$t\!=\!0$, $Q(0\!\mid\!x_{0})\!=\!1$, which results from the initial condition
$p(x,0\!\mid\!x_{0})\!=\!\delta(x-x_{0})$ for the probability distribution and
integration of the Dirac delta function with respect to $x$. The last property
emerges from our conviction that the diffusing particle will eventually end up
at the target $x_{\mathrm{f}}$ after a very long time. In other words, if
$t\!\to\!\infty$ then $Q(\infty\!\mid\!x_{0})\!=\!0$. The second and third
properties of the survival probability make the first-passage time distribution
normalized to unity, namely $\int_{0}^{\infty}F(t\!\mid\!x_{0})\,\mathrm{d}t\!=
\!1$. This means the particle is sure to hit the target for the first time,
although the mean time, by which such an event occurs, need not be always
finite. In addition, inserting Eq.~(\ref{eq_6}) into Eq.~(\ref{eq_5}) and
performing integration by parts, we readily show that the mean first-passage
time
\begin{equation}
\mathcal{T}(x_{0}\!\to\!x_{\mathrm{f}})=\int_{0}^{\infty}Q(t\!\mid\!x_{0})\,\mathrm{d}t.
\label{eq_7}
\end{equation}

Let us now establish the duo of coupled equations that make it possible to combine
the probability distribution, e.g. such as in Eq.~(\ref{eq_4}), with the survival
probability. The first equation we have already met in Eq.~(\ref{eq_6}), 
constitutes the relationship between the survival probability $Q(t\!\mid\!x_{0})$
and the first-passage time distribution $F(t\!\mid\!x_{0})$. The second equation
relates the first-passage time distribution with the probability distribution, and
has the form of the integral equation~\cite{Krap2010}
\begin{equation}
p(0,t\!\mid\!x_{0})=\int_{0}^{t}F(\tau\!\mid\!x_{0})\,p(0,t-\tau\!\mid\!0)
\,\mathrm{d}\tau.
\label{eq_8}
\end{equation}
This equation defines the probability distribution or, more precisely, the
propagator from $x\!=\!x_{0}$ to the target at $x_{\text{f}}\!=\!0$ for any
stochastic dynamics as an integral over the first time to reach the position $0$
at a time $\tau\!\leqslant\!t$ followed by a loop from the spatiotemporal
coordinate $(0,\tau)$ to $(0,t)$ in the remaining time $t\!-\!\tau$. The target
$x_{\mathrm{f}}\!=\!0$ may correspond, for instance, to the minimum of the
harmonic potential, which is crucial for the Ornstein-Uhlenbeck process. Note
that the integral expression in the above equation is a time convolution of
two distribution functions, thus a price we must pay to determine the
first-passage time distribution $F(t\!\mid\!x_{0})$ involves the use of the
Laplace transformation. The convolution theorem states that the Laplace
transformation, defined as $\tilde{f}(s)\!=\!\mathcal{L}[f(t);t]\!:=\!
\int_{0}^{\infty}f(t)\,\mathrm{e}^{-st}\mathrm{d}t$, of the convolution
$f(t)\!\ast\!g(t)\!:=\!\int_{0}^{t}f(\tau)g(t-\tau)\mathrm{d}\tau$ of two
integrable functions $f(t)$ and $g(t)$ is the product of their Laplace 
transforms, i.e. $\mathcal{L}[f(t)\!\ast\!g(t);t]\!=\!\tilde{f}(s)\tilde{g}
(s)$~\cite{Schi1999}. Therefore, we can convert Eq.~(\ref{eq_8}) into the
algebraic form
\begin{equation}
\tilde{F}(s\!\mid\!x_{0})=\frac{\tilde{p}(0,s\!\mid\!x_{0})}
{\tilde{p}(0,s\!\mid\!0)}.
\label{eq_9}
\end{equation}
In turn, performing the Laplace transformation of Eq.~(\ref{eq_6}) yields
\begin{equation}
\tilde{F}(s\!\mid\!x_{0})=1-s\,\tilde{Q}(s\!\mid\!x_{0}),
\label{eq_10}
\end{equation}
where the Laplace transform $\mathcal{L}\!\left[\frac{\mathrm{d}}{\mathrm{d}t}
Q(t\!\mid\!x_{0});t\right]\!=\!s\,\tilde{Q}(s\!\mid\!x_{0})-Q(0\!\mid\!x_{0})\!
=\!s\,\tilde{Q}(s\!\mid\!x_{0})-1$ has been performed. The combination of the
last two expressions implies a direct relationship between the survival
probability and the ratio of probability distributions (return probability
distribution in the denominator) in the Laplace domain:
\begin{equation}
\tilde{Q}(s\!\mid\!x_{0})=\frac{1}{s}\left[1-\frac{\tilde{p}(0,s\!\mid\!x_{0})}
{\tilde{p}(0,s\!\mid\!0)}\right].
\label{eq_11}
\end{equation}
Armed with the above equation and Eq.~(\ref{eq_7}), we can determine the mean
first-passage time from the initial position $x_{0}$ to the target point
$x_{\mathrm{f}}\!=\!0$ as follows:
\begin{equation}
\mathcal{T}(x_{0}\!\to\!0)=\lim_{s\to0}\int_{0}^{\infty}\!
Q(t\!\mid\!x_{0})\,\mathrm{e}^{-st}\,\mathrm{d}t=\lim_{s\to0}
\mathcal{L}[Q(t\!\mid\!x_{0});s]=\lim_{s\to0}\tilde{Q}(s\!\mid\!x_{0}).
\label{eq_12}
\end{equation}
On the other hand, performing the inverse Laplace transformation of 
Eq.~(\ref{eq_11}), which usually is not a trivial operation, allows one to find
the survival probability and hence the first-passage time distribution from 
Eq.~(\ref{eq_6}) in real space.

There exists an independent method worth mentioning here because of its common
use in finding the survival probability. This method is rooted in the backward
Fokker-Planck equation that in case of diffusion occurring in the
confining potential $V(x)$ has the following structure~\cite{Gard2004}:
\begin{equation}
\frac{\partial}{\partial t}Q(t\!\mid\!x)=-\frac{\mathrm{d}V(x)}{\mathrm{d}x}
\frac{\partial}{\partial x}Q(t\!\mid\!x)+D\frac{\partial^{2}}{\partial x^{2}}
Q(t\!\mid\!x).
\label{eq_13}
\end{equation}
We can solve the above partial differential equation in the Laplace domain
assuming the initial condition $Q(0\!\mid\!x_{0})\!=\!1$ at position
$x\!=\!x_{0}$, the absorbing boundary condition $Q(t\!\mid\!x_{\text{f}})\!=\!0$
at position $x\!=\!x_{\text{f}}$ and the additional condition $Q(t\!\mid\!
\infty)\!=\!1$. The last condition means that a particle is sure to survive
until time $t$, being infinity distant from the absorbing point. Having the
solution of Eq.~(\ref{eq_13}) at our disposal, we can then substitute it into
Eq.~(\ref{eq_10}) to obtain the Laplace transform of the first-passage time
distribution $\tilde{F}(s\!\mid\!x_{0})$ or directly into Eq.~(\ref{eq_12})
to determine the mean first-passage time. Furthermore, using Eqs.~(\ref{eq_5})
and (\ref{eq_9}) makes it possible to show that
\begin{align}
\mathcal{T}(x_{0}\!\to\!0)&=\int_{0}^{\infty}\!tF(t\!\mid\!x_{0})\,
\mathrm{d}t=\lim_{s\to0}\int_{0}^{\infty}\!tF(t\!\mid\!x_{0})\,\mathrm{e}^{-st}\,
\mathrm{d}t=-\lim_{s\to0}\frac{\partial}{\partial s}\int_{0}^{\infty}\!
F(t\!\mid\!x_{0})\,\mathrm{e}^{-st}\,\mathrm{d}t\nonumber\\
&=-\lim_{s\to0}\frac{\partial}{\partial s}\mathcal{L}[F(t\!\mid\!x_{0});s]
=-\lim_{s\to0}\frac{\partial}{\partial s}\tilde{F}(s\!\mid\!x_{0}).
\label{eq_14}
\end{align}
We will utilize the above formula to find the exact result for the mean
first-passage time downhill of the harmonic potential.

The backward Fokker-Planck equation (\ref{eq_13}) for the survival probability
with appropriate boundary conditions constitutes a prototype of the partial
differential equation for the very mean first-passage time. In fact, taking the
integral with respect to the time of both sides of Eq.~(\ref{eq_13}) in accordance
with Eq.~(\ref{eq_7}), or Laplace transform of this equation, followed by the
limit as in the last component of Eq.~(\ref{eq_12}), we readily show that
\begin{equation}
D\frac{\mathrm{d}^{2}\mathcal{T}(x)}{\mathrm{d}x^{2}}-\frac{\mathrm{d}V(x)}
{\mathrm{d}x}\frac{\mathrm{d}\mathcal{T}(x)}{\mathrm{d}x}=-1.
\label{eq_15}
\end{equation}
To find an unambiguous solution to this differential equation with a given
potential $V(x)$, we need to extend it with mixed Dirichlet-von Neumann boundary
conditions~\cite{Greb2016}. For a particle already at the absorbing target point
$x\!=\!x_{\text{f}}$, it is clear that $\mathcal{T}(x_{\text{f}})\!=\!0$, while
at the reflecting point $r$, located in the potential in such a way that
$r\!>\!x_{\text{f}}$ or $r\!<\! x_{\text{f}}$, the derivative of the mean
first-passage time at $x\!=\!r$ is $\frac{\mathrm{d}\mathcal{T}(x)}{\mathrm{d}x}\!
\Big|_{x\!=\!r}\!=\!0$. We will utilize this method as the next argument in
proving the rigorous solution to the mean first-passage time downward and also
upward of the harmonic potential.

We will also argue that the same results can be obtained by taking advantage of
an alternative well-known formula
\begin{equation}
\mathcal{T}_{\scriptscriptstyle{\swarrow}}(x_{0}\!\to\!0)=\frac{1}{D}
\int_{0}^{x_{0}}\!\mathrm{d}y\,\exp\!\left[\frac{V(y)}{D}\right]\!
\int_{y}^{\infty}\!\exp\!\left[-\frac{V(z)}{D}\right]\mathrm{d}z,
\label{eq_16}
\end{equation}
which actually emerges from a direct integration of Eq.~(\ref{eq_15}). Here, we
assume that a particle starting from the initial position $x_{0}\!>\!0$ diffuses
downhill of the potential $V(x)$ to reach the target at $x_{\mathrm{f}}\!=\!0$,
while the reflecting barrier $r\!>\!x_{0}\!>\!0$ has been pushed to infinity
~\cite{Gard2004}. In this way, using three independent methods, we will make sure
that our final result turns out to be correct. We can also write down the second
variant of Eq.~(\ref{eq_16}) which, in contrast to the formula in
Eq.~(\ref{eq_14}), will allow us to determine the mean first-passage time upward
of the harmonic potential. Thus, if the particle diffuses from the initial
position at $x\!=\!0$ to the target point at $x_{\text{f}}\!=\!x_{0}\!>\!0$
(see Fig.~1), then
\begin{equation}
\mathcal{T}_{\scriptscriptstyle{\nearrow}}(0\!\to\!x_{0})=\frac{1}{D}\int_{0}
^{x_{0}}\!\mathrm{d}y\,\exp\!\left[\frac{V(y)}{D}\right]\!\int_{-\infty}^{y}\!
\exp\!\left[-\frac{V(z)}{D}\right]\mathrm{d}z,
\label{eq_17}
\end{equation}
where we have assumed the reflecting barrier $r\!<\!0\!<\!x_{0}$ to be at minus
infinity.

\section*{3.~Mean first-passage time for diffusion downward of harmonic
         \newline potential}

According to Eq.~(\ref{eq_4}), the probability distribution of finding a diffusing
particle in the minimum $x\!=\!0$ of the harmonic potential at time $t\!>\!0$,
given that it was initially positioned at $x_{0}\!>\!0$ is
\begin{equation}
p(0,t\!\mid\!x_{0})=\sqrt{\frac{\alpha}{2\pi D\!\left(1-\mathrm{e}^{-2\alpha t}
\right)}}\exp\!\left[-\frac{\alpha\,x^{2}_{0}\,\mathrm{e}^{-2\alpha t}}{2D\!
\left(1-\mathrm{e}^{-2\alpha t}\right)}\right].
\label{eq_18}
\end{equation}
To proceed, the Laplace transform of the above distribution function has to be
performed. For this purpose, we take advantage of the following formula:
\begin{equation}
\int_{0}^{\infty}\left(\mathrm{e}^{\tau}-1\right)^{\nu-1}\exp\!\left(-\frac{z}
{\mathrm{e}^{\tau}-1}-\mu\tau\right)\mathrm{d}\tau=\Gamma(\mu-\nu+1)\,
\mathrm{e}^{z/2}\,z^{(\nu-1)/2}\,W_{\frac{\nu-2\mu-1}{2},\frac{\nu}{2}}(z),
\label{eq_19}
\end{equation}
in which $W_{\gamma,\beta}(z)\!=\!z^{\beta+1/2}\exp\!\left(-\frac{z}{2}\right)
U\!\left(\beta-\gamma+\frac{1}{2},2\beta+1,z\right)$ is the Whittaker
hypergeometric function defined by the Tricomi confluent hypergeometric function
$U(a,b,z)$~\cite{Grad2007}. The former function satisfies the identity
$W_{\gamma,\beta}(z)\!=\!W_{\gamma,-\beta}(z)$ arising from the functional
identity of the latter, that is, $U(a,b,z)\!=\!z^{1-b}U(a-b+1,2-b,z)$. Taking 
into account all these properties along with Eq.~(\ref{eq_19}), we find that the
Laplace transform of the probability distribution in Eq.~(\ref{eq_18}) reads
\begin{equation}
\tilde{p}(0,s\!\mid\!x_{0})=\frac{\Gamma\!\left(\frac{s}{2\alpha}\right)}
{\sqrt{8\pi D\,\alpha}}\,U\!\left(\frac{s}{2\alpha},\frac{1}{2},
\frac{\alpha\,x_{0}^{2}}{2D}\right).
\label{eq_20}
\end{equation}
On the other hand, if the initial position $x_{0}\!>\!0$ coincides with the target
point at the minimum $x\!=\!0$ of the harmonic potential, then Eq.~(\ref{eq_18})
yields
\begin{equation}
p(0,t\!\mid\!0)=\sqrt{\frac{\alpha}{2\pi D\!\left(1-\mathrm{e}^{-2\alpha t}
\right)}}.
\label{eq_21}
\end{equation}
To determine the Laplace transform of this return probability distribution, we use
the integral $\int_{0}^{\infty}\left(1-\mathrm{e}^{-\tau/\lambda}\right)^{\nu-1}
\linebreak\mathrm{e}^{-\mu\tau}\mathrm{d}\tau\!=\!\lambda\,\mathrm{B}(\lambda
\mu,\nu)$, where $\mathrm{B}(\mu,\nu)\!=\!\frac{\Gamma(\mu)\Gamma(\nu)}{\Gamma(
\mu+\nu)}$ is the Euler beta function. Consequently, we find from
Eq.~(\ref{eq_21}) that
\begin{equation}
\tilde{p}(0,s\!\mid\!0)=\frac{\Gamma\!\left(\frac{s}{2\alpha}\right)}
{\sqrt{8D\alpha}\,\Gamma\!\left(\frac{s}{2\alpha}+\frac{1}{2}\right)}.
\label{eq_22}
\end{equation}
Given Eqs.~(\ref{eq_20}) and (\ref{eq_22}) enable us to perform two independent
operations taking into account Eq.~(\ref{eq_7}), see also Eq.~(\ref{eq_12}), or
Eq.~(\ref{eq_13}). According to the first scenario, the Laplace transform of the
survival probability in Eq.~(\ref{eq_11}) is
\begin{equation}
\tilde{Q}(s\!\mid\!x_{0})=\frac{1}{s}-\frac{\Gamma\!\left(\frac{s}{2\alpha}\right)
\Gamma\!\left(\frac{s}{2\alpha}+\frac{1}{2}\right)}{2\sqrt{\pi}\alpha\,
\Gamma\!\left(\frac{s}{2\alpha}+1\right)}\,U\!\left(\frac{s}{2\alpha},\frac{1}{2},
\frac{\alpha\,x_{0}^{2}}{2D}\right).
\label{eq_23}
\end{equation}
To execute the inverse Laplace transformation of the above expression and return
to the time domain $\tau$, we take advantage of the following relation
~\cite{Ober1973}:
\begin{equation}
\mathcal{L}^{-1}\!\left[\frac{\Gamma\!\left(\frac{1}{2}+\beta+s\right)\Gamma\!
\left(\frac{1}{2}-\beta+s\right)}{\Gamma\!(1-\gamma+s)}W_{-s,\beta}(z);s\right]=
\frac{\mathrm{e}^{-z/2}}{\left(1-\mathrm{e}^{-\tau}\right)^{\gamma}}\exp\!\left[-
\frac{z}{2\left(\mathrm{e}^{\tau}-1\right)}\right]W_{\gamma,\beta}\!\left(\frac{z}
{\mathrm{e}^{\tau}-1}\right).
\label{eq_24}
\end{equation}
Setting in this formula $\gamma\!=\!-\frac{1}{4}$, $\beta\!=\!\frac{1}{4}$, then
changing $s\!\to\!as\!-\!b$ and using the general property of the inverse Laplace
transform $\mathcal{L}^{-1}[f(as-b);s]=\frac{1}{a}\mathrm{e}^{bt/a}f\!
\left(\frac{t}{a}\right)$ for $a\!=\!\frac{1}{2\alpha}$ and $b\!=\!\frac{1}{4}$,
as well as considering the aforementioned relation between the Whittaker and
Tricomi hyperbolic functions, we obtain that
\begin{equation}
\mathcal{L}^{-1}\!\left[\frac{\Gamma\!\left(\frac{s}{2\alpha}\right)\Gamma\!\left(
\frac{s}{2\alpha}+\frac{1}{2}\right)}{\Gamma\!\left(\frac{s}{2\alpha}+1\right)}
U\!\left(\frac{s}{2\alpha},\frac{1}{2},z\right);s\right]=2\alpha\exp\!\left[
-\frac{z}{\left(\mathrm{e}^{2\alpha t}-1\right)}\right]U\!\left(\frac{1}{2},
\frac{1}{2},\frac{z}{\left(\mathrm{e}^{2\alpha t}-1\right)}\right).
\label{eq_25}
\end{equation}
A direct application of this inverse Laplace transformation to Eq.~(\ref{eq_23})
results in the survival probability
\begin{equation}
Q(t\!\mid\!x_{0})=1-\frac{1}{\sqrt{\pi}}\exp\!\left[-\frac{\alpha\,x_{0}^{2}}
{2D\left(\mathrm{e}^{2\alpha t}-1\right)}\right]U\!\left(\frac{1}{2},
\frac{1}{2},\frac{\alpha\,x_{0}^{2}}{2D\left(\mathrm{e}^{2\alpha t}-1\right)}
\right).
\label{eq_26}
\end{equation}
We can further simplify this formula knowing that
\begin{equation}
U\!\left(\frac{1}{2},\frac{1}{2},z\right)=\sqrt{\pi}\,\mathrm{e}^{z}\mathrm{erfc}
(\sqrt{z}),
\label{eq_27}
\end{equation}
where $\mathrm{erfc}(z)\!=\!1-\mathrm{erf}(z)$ is the complementary error function
and $\mathrm{erf}(z)$ stands for the error function~\cite{Bryc2008}. In this way,
the final formula for the survival probability simplifies to the following form:
\begin{equation}
Q(t\!\mid\!x_{0})=\mathrm{erf}\!\left(\frac{\sqrt{\alpha}\,\lvert x_{0}\rvert}{
\sqrt{2D\left(\mathrm{e}^{2\alpha t}-1\right)}}\right),
\label{eq_28}
\end{equation}
whereas the mean first-passage time in Eq.~(\ref{eq_7}) from the initial position
$x_{0}\!>\!0$ downhill of the harmonic potential to its minimum at $x\!=\!0$ is
\begin{equation}
\mathcal{T}_{\scriptscriptstyle{\swarrow}}(x_{0}\!\to\!0)=\int_{0}^{\infty}
\mathrm{erf}\!\left(\frac{\sqrt{\alpha}\,\lvert x_{0}\rvert}
{\sqrt{2D\left(\mathrm{e}^{2\alpha t}-1\right)}}\right)\mathrm{d}t.
\label{eq_29}
\end{equation}
Unquestionably, the exact calculation of the above integral is a formidable task.
In this situation, we are forced to resort to the procedure of numerical
integration in order to compute the mean first-passage time in Eq.~(\ref{eq_29}).
However, in the close vicinity of the target point $x\!=\!0$, that is, when the
initial position $x_{0}$ goes to $0$, we can approximate the error function
$\mathrm{erf}(z)\!\propto\!\frac{2z}{\sqrt{\pi}}$ for $z\!\to\!0$ and next utilize
the integral $\int_{0}^{\infty}\left(1-\mathrm{e}^{-\tau/\lambda}\right)^{\nu-1}
\mathrm{e}^{-\mu\tau}\mathrm{d}\tau\!=\!\lambda\,\mathrm{B}(\lambda\mu,\nu)$ to
show that
\begin{equation}
\mathcal{T}_{\scriptscriptstyle{\swarrow}}(x_{0}\!\to\!0)\simeq\sqrt{\frac{\pi}
{2D\alpha}}x_{0}.
\label{eq_30}
\end{equation}
The opposite extreme approximation of the integral in Eq.~(\ref{eq_29}), namely
for $\lvert x_{0}\rvert\!\to\!\infty$, is rather hard to study due to the
asymptotic representation of the error function.

Let us now realize the second scenario inserting Eqs.~(\ref{eq_20}) and
(\ref{eq_22}) into Eq.~(\ref{eq_9}). Then, the Laplace transform of the
first-passage time distribution is
\begin{equation}
\tilde{F}(s\!\mid\!x_{0})=\frac{1}{\sqrt{\pi}}\Gamma\!\left(\frac{s}{2\alpha}+
\frac{1}{2}\right)U\!\left(\frac{s}{2\alpha},\frac{1}{2},\frac{\alpha\,x^{2}_{0}}
{2D}\right).
\label{eq_31}
\end{equation}
To effectively use Eq.~(\ref{eq_14}) in order to determine the mean first-passage
time downhill of the harmonic potential, we need to begin with a calculation of
the first derivative of this function with respect to the Laplace variable $s$.
The function in Eq.~(\ref{eq_31}) is the product of the Euler gamma function
$\Gamma(z)$ and the Tricomi confluent hypergeometric function $U(a,b,z)$. So, we
have in general that
\begin{equation}
\frac{\partial}{\partial z}\Gamma\!\left(z+\frac{1}{2}\right)=\psi\!\left(z+
\frac{1}{2}\right)\Gamma\!\left(z+\frac{1}{2}\right),
\label{eq_32}
\end{equation}
where $\psi(z)\!=\!\frac{\mathrm{d}\log\Gamma(z)}{\mathrm{d}z}$ is a digamma
function, while in the case of the second function and its derivative over the
first parameter $a$, we get the following result:
\begin{align}
\frac{\partial}{\partial a}U(a,b,z)&=-\frac{\Gamma(1-b)\,\psi(a-b+1)}
{\Gamma(a-b+1)}\,\pFq{1}{1}(a;b;z)-\frac{\Gamma(b-1)\,\psi(a)\,z^{1-b}}
{\Gamma(a)}\,\pFq{1}{1}(a-b+1;2-b;z)\nonumber\\
&-\frac{\Gamma(-b)\,z}{\Gamma(a-b+1)}\,
F_{\hspace{0.6mm}2:0;1}^{1:1;2}\!\left[\left.
\begin{matrix}
a\!+\!1\\
2,b\!+\!1
\end{matrix}
\begin{matrix}
\,:\,\\
\,:\,
\end{matrix}
\begin{matrix}
1\\
-
\end{matrix}
\begin{matrix}
\,;\,\\
\,;\,
\end{matrix}
\begin{matrix}
1,a\\
a\!+\!1
\end{matrix}
\right| z, z
\right]-
\frac{\Gamma(b-2)\,z^{2-b}}{\Gamma(a)}\,
F_{\hspace{0.6mm}2:0;1}^{1:1;2}\!\left[\left.
\begin{matrix}
a\!-\!b\!+\!2\\
2,3\!-\!b
\end{matrix}
\begin{matrix}
\,:\,\\
\,:\,
\end{matrix}
\begin{matrix}
1\\
-
\end{matrix}
\begin{matrix}
\,;\,\\
\,;\,
\end{matrix}
\begin{matrix}
1,a\!-\!b\!+\!1\\
a\!-\!b\!+\!2
\end{matrix}
\right| z, z
\right]\nonumber\\
\label{eq_33}
\end{align}
for $b\!\notin\!\mathbb{Z}$, where $\pFq{1}{1}(a;b;z)$ is the Kummer confluent
hypergeometric function~\cite{Duve2024} and the Kamp\'e de F\'eriet hypergeometric
function in two variables $x$, $y$ is defined by a double power 
series~\cite{Cvij2010,Choi2021,Pari2022}:
\begin{equation}
F_{\hspace{0.6mm}\lambda:\mu;\nu}^{\alpha:\beta;\gamma}\!\left[\left.
\begin{matrix}
(a_{\alpha})\\
(b_{\lambda})
\end{matrix}
\begin{matrix}
:\\
:
\end{matrix}
\begin{matrix}
(c_{\beta})\\
(d_{\mu})
\end{matrix}
\begin{matrix}
;\\
;
\end{matrix}
\begin{matrix}
(f_{\gamma})\\
(g_{\nu})
\end{matrix}
\right| x, y
\right]
=\sum_{m=0}^{\infty}\sum_{n=0}^{\infty}
\frac{((a))_{m+n} ((c))_{m} ((f))_{n}  
}{((b))_{m+n} ((d))_{m} ((g))_{n}} 
\frac{x^{m}}{m!}\frac{y^{n}}{n!}.
\label{eq_34}
\end{equation}
Here, $(a)_{n}\!:=\!\Gamma(a+n)/\Gamma(a)$ means the Pochhammer symbol for an
integer $n$, the symbol $(a_{m})$ denotes the sequence $(a_{1},\dots,a_{m})$ and
the product of $m$ Pochhammer symbols $((a_{m}))$ is determined by $((a_{m}))_{n}
\!:=\!(a_{1})_{n}\cdots(a_{m})_{n}$. In addition, the empty product for $m\!=\!0$
reduces to unity.

Let us now prepare the base for further calculations. At first, we want to show
a direct relationship between the Kamp\'e de F\'eriet hypergeometric function and
the generalized hypergeometric function, which the primary definition is as
follows:
\begin{equation}
\pFq{2}{2}(a_{1},a_{2};b_{1},b_{2};z)=\sum_{m=0}^{\infty}\frac{(a_{1})_{m}
(a_{2})_{m}}{(b_{1})_{m}(b_{2})_{m}}\frac{z^{m}}{m!}.
\label{eq_35}
\end{equation}
Indeed, setting in Eq.~(\ref{eq_34}) $\alpha\!=\!\beta\!=\!\nu\!=\!1$,
$\gamma\!=\!\lambda\!=\!2$, $\mu\!=\!0$ and $x\!=\!y\!=\!z$ implies that
\begin{align}
F_{\hspace{0.6mm}2:0;1}^{1:1;2}\!\left[\left.
\begin{matrix}
1\\
2,\frac{\scriptscriptstyle{3}}{\scriptscriptstyle{2}}
\end{matrix}
\begin{matrix}
\,:\,\\
\,:\,
\end{matrix}
\begin{matrix}
1\\
-
\hspace{1mm}
\end{matrix}
\begin{matrix}
\,;\,\\
\,;\,
\end{matrix}
\begin{matrix}
1,0\\
1
\end{matrix}
\right| z, z
\right]
&=\sum_{m=0}^{\infty}\sum_{n=0}^{\infty}
\frac{(1)_{m+n}(1)_{m}(1)_{n}(0)_{n}}{(2)_{m+n}\left(\frac{\scriptscriptstyle3}
{\scriptscriptstyle2}\right)_{m+n}(1)_{n}}\frac{z^{m+n}}{m!n!}\nonumber\\
&=\sum_{m=0}^{\infty}\frac{(1)_{m}(1)_{m}}{(2)_{m}\!
\left(\frac{\scriptscriptstyle3}{\scriptscriptstyle2}\right)_{m}}\frac{z^{m}}{m!},
\label{eq_36}
\end{align}
where the second line results from the summation over index $n$ and two special
values of the Pochhammer symbol, such as $(0)_{0}\!=\!1$ and $(0)_{n}\!=\!0$ for
$n\!\in\!\mathbb{N}$. On the other hand, assuming $a_{1}\!=\!a_{2}\!=\!1$,
$b_{1}\!=\!\frac{3}{2}$ and $b_{2}\!=\!2$, we readily rewrite Eq.~(\ref{eq_35})
as
\begin{equation}
\pFq{2}{2}\!\left(1,1;\frac{3}{2},2;z\right)=\sum_{m=0}^{\infty}\frac{(1)_{m}
(1)_{m}}{(2)_{m}\!\left(\frac{\scriptscriptstyle3}
{\scriptscriptstyle2}\right)_{m}}\frac{z^{m}}{m!},
\label{eq_37}
\end{equation}
while a comparison of the right-hand sides of the last two formulae implicates that
\begin{equation}
F_{\hspace{0.6mm}2:0;1}^{1:1;2}\!\left[\left.
\begin{matrix}
1\\
2,\frac{\scriptscriptstyle{3}}{\scriptscriptstyle{2}}
\end{matrix}
\begin{matrix}
\,:\,\\
\,:\,
\end{matrix}
\begin{matrix}
1\\
-
\hspace{1mm}
\end{matrix}
\begin{matrix}
\,;\,\\
\,;\,
\end{matrix}
\begin{matrix}
1,0\\
1
\end{matrix}
\right| z, z
\right]=\pFq{2}{2}\!\left(1,1;\frac{3}{2},2;z\right).
\label{eq_38}
\end{equation}
At second, we need to know that $U(0,b,z)\!=\!1$ and $\pFq{1}{1}(0,b,z)\!=\!1$
for any $b$ and $z$, whereas $\Gamma\!\left(\frac{1}{2}\right)\!=\!\sqrt{\pi}$ and
$\Gamma\!\left(-\frac{1}{2}\right)\!=\!-2\sqrt{\pi}$. Moreover, the two relations,
$\lim_{z\to0}\frac{\psi(z)}{\Gamma(z)}\!=\!-1$ and $\lim_{z\to0}\frac{\Gamma\!
\left(z+\frac{1}{2}\right)}{\Gamma(z)}\!=\!0$, hold.
\begin{figure}[t]
\centering
\includegraphics[scale=0.3]{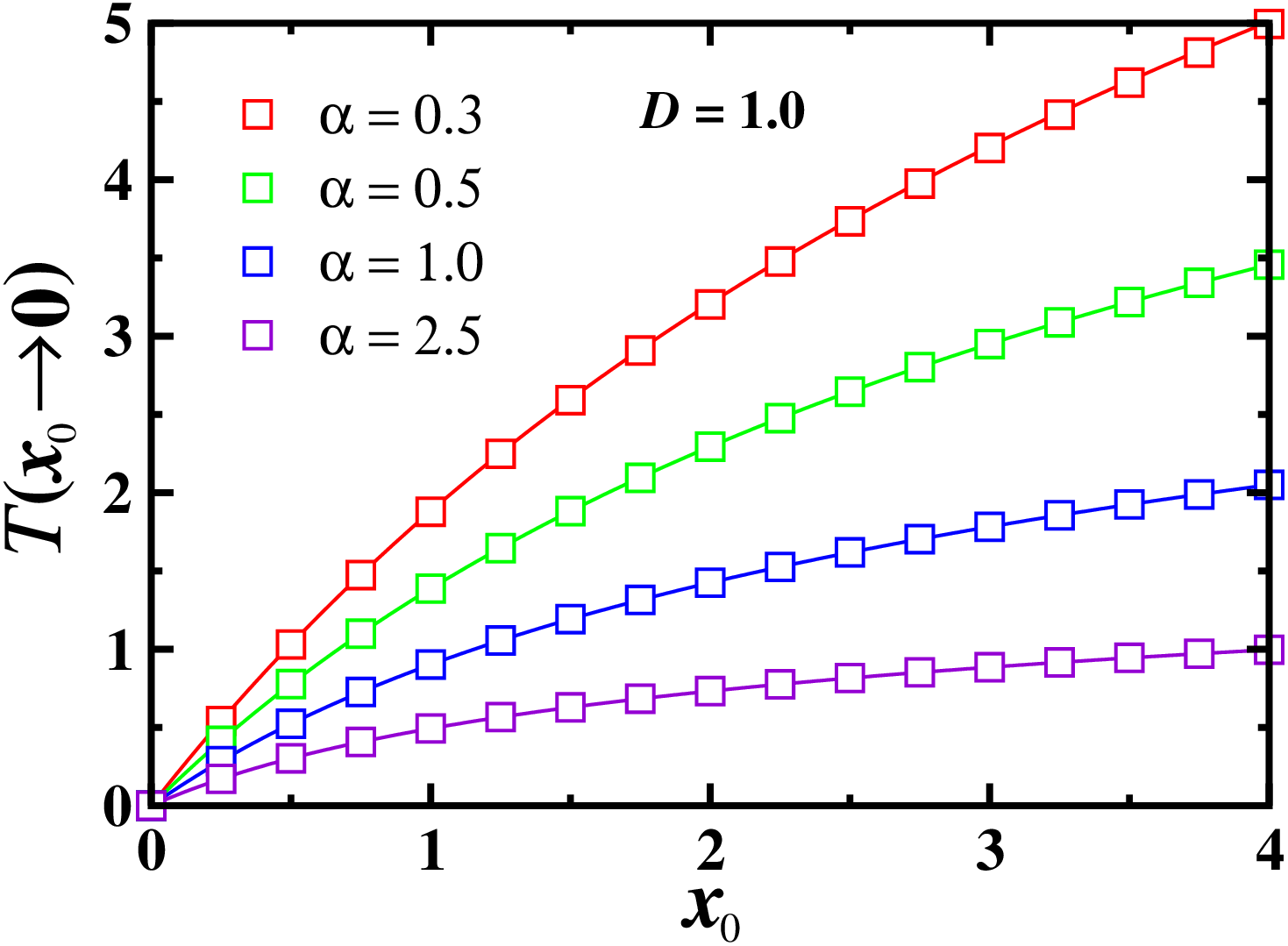}
\caption{Excellent compatibility of the analytical formula for the mean
first-passage time (solid lines) given by Eq.~(\ref{eq_41}) and numerical data
(squared points) obtained from integration performed in Eq.~(\ref{eq_29}). A few
values of the parameter $\alpha$ have been established and the diffusion constant
$D\!=\!1.0$ has been assumed.}
\label{fig2}
\end{figure}

Therefore, taking into account Eqs.~(\ref{eq_32}) and (\ref{eq_33}) along with
Eq.~(\ref{eq_38}) and all the above properties, we are able to affirm that
\begin{equation}
\lim_{s\to0}\,U\!\left(\frac{s}{2\alpha},\frac{1}{2},\frac{\alpha x_{0}^{2}}
{2D}\right)\frac{\partial}{\partial s}\,\Gamma\!\left(\frac{s}{2\alpha}+
\frac{1}{2}\right)=\frac{\sqrt{\pi}}{2\alpha}\psi\!\left(\frac{1}{2}\right),
\label{eq_39}
\end{equation}
where the specific value of the digamma function $\psi\!\left(\frac{1}{2}\right)
\!=\!-\log4-\gamma$ with the Euler-Mascheroni constant $\gamma\!\approx\!0.5772$,
and
\begin{align}
\lim_{s\to0}\,\Gamma\!\left(\frac{s}{2\alpha}+\frac{1}{2}\right)\frac{\partial}
{\partial s}\,U\!\left(\frac{s}{2\alpha},\frac{1}{2},
\frac{\alpha x_{0}^{2}}{2D}\right)&=-\frac{\sqrt{\pi}}{2\alpha}\psi\!\left(\frac{1}
{2}\right)-\frac{\pi x_{0}}{\sqrt{2D\alpha}}\,\pFq{1}{1}\!\left(\frac{1}{2};
\frac{3}{2};\frac{\alpha x_{0}^{2}}{2D}\right)+\frac{\sqrt{\pi}x_{0}^{2}}{2D}\,
\pFq{2}{2}\!\left(1,1;\frac{3}{2},2;\frac{\alpha x_{0}^{2}}{2D}\right).\nonumber\\
\label{eq_40}
\end{align}
Calculating the first derivative of the Laplace transform of the first-passage
time distribution in Eq.~(\ref{eq_31}), next inserting it into Eq.~(\ref{eq_14})
and taking advantage of Eqs.~(\ref{eq_39}) and (\ref{eq_40}) results in the final
expression for the mean first-passage time from the initial position $x_{0}\!>\!
0$ downhill of the harmonic potential to its minimum at $x\!=\!0$. The exact 
formula has the following form:
\begin{equation}
\mathcal{T}_{\scriptscriptstyle{\swarrow}}(x_{0}\!\to\!0)=\sqrt{\frac{\pi x_{0}
^{2}}{2D\alpha}}\,\pFq{1}{1}\!\left(\frac{1}{2};\frac{3}{2};\frac{\alpha x_{0}
^{2}}{2D}\right)-\frac{x_{0}^{2}}{2D}\,\pFq{2}{2}\!\left(1,1;\frac{3}{2},2;
\frac{\alpha x_{0}^{2}}{2D}\right).
\label{eq_41}
\end{equation}
We show in Fig.~\ref{fig2} excellent agreement between the above analytical
result (solid lines) and the data (squared points) obtained by numerical
integration of Eq.~(\ref{eq_29}). A number of different values of the parameter
$\alpha$ have been selected, which decide on the strength of the harmonic
potential, and the diffusion coefficient $D\!=\!1$ has been assumed.

Prior to commenting on the central result given by Eq.~(\ref{eq_41}), let us
first verify its correctness utilizing a completely different method which 
culminates in Eq.~(\ref{eq_16}). Thus, in the special case of the harmonic
potential $V(x)\!=\!\frac{1}{2}\alpha x^{2}$, we take advantage of the integral
$\int_{u}^{\infty}\exp\!\left(-\frac{z^{2}}{4\gamma}\right)\mathrm{d}z\!=\!
\sqrt{\pi\gamma}\,\mathrm{erfc}\!\left(\frac{u}{2\sqrt{\gamma}}\right)$ to find
that
\begin{equation}
\mathcal{T}_{\scriptscriptstyle{\swarrow}}(x_{0}\!\to\!0)=\sqrt{\frac{\pi}
{2D\alpha}}\int_{0}^{x_{0}}\exp\!\left(\frac{\alpha y^{2}}{2D}\right)
\mathrm{erfc}\!\left(\sqrt{\frac{\alpha}{2D}}y\right)\mathrm{d}y.
\label{eq_42}
\end{equation}
Interestingly, the remaining integral can be also strictly performed in two ways.
Hereafter, we show these calculations separately. The first approach is to replace
the integrand in Eq.~(\ref{eq_42}) by the one-parametric Mittag-Leffler function
$E_{\theta}(w)\!=\!\sum_{n=0}^{\infty}\frac{w^{n}}{\Gamma(\theta n+1)}$. This is
possible because for the parameter $\theta\!=\!\frac{1}{2}$, $E_{1/2}
(-\sqrt{w})\!=\!\exp(w)\,\mathrm{erfc}(\sqrt{w})$~\cite{Gore2014}. On the other
hand, the half-parameter Mittag-Leffler function
\begin{equation}
E_{1/2}(z)=\pFq{0}{0}\!\left(;\,;z^{2}\right)+\frac{2z}{\sqrt{\pi}}\pFq{1}{1}\!
\left(1;\frac{3}{2};z^{2}\right),
\label{eq_43}
\end{equation}
where $\pFq{0}{0}(;\,;z)$ is the generalized hypergeometric function and
$\pFq{1}{1}(a;b;z)$ corresponds to the Kummer confluent hyperbolic function.
Inserting $z\!=\!-\sqrt{w}$ into Eq.~(\ref{eq_43}) and correspondingly $w\!=\!
\frac{\alpha y^{2}}{2D}$, we get from Eq.~(\ref{eq_42}) that
\begin{equation}
\mathcal{T}_{\scriptscriptstyle{\swarrow}}(x_{0}\!\to\!0)=\sqrt{\frac{\pi}
{2D\alpha}}\int_{0}^{x_{0}}\pFq{0}{0}\!\left(;\,;\frac{\alpha\,y^{2}}
{2D}\right)\mathrm{d}y-\frac{1}{D}\int_{0}^{x_{0}}y\,\pFq{1}{1}\!\left(1;\frac{3}
{2};\frac{\alpha\,y^{2}}{2D}\right)\mathrm{d}y.
\label{eq_44}
\end{equation}
Both integrals in the above expression are precisely solvable. The first integral
\begin{equation}
\int_{0}^{u}\pFq{0}{0}\!\left(;\,;az^{2}\right)\mathrm{d}z=\sqrt{\frac{\pi}{4a}}
\mathrm{erfi}\left(\sqrt{a}u\right),
\label{eq_45}
\end{equation}
where $\mathrm{erfi}(z)\!=\!\frac{2z}{\sqrt{\pi}}\pFq{1}{1}\!\left(\frac{1}{2};
\frac{3}{2};z^{2}\right)$ is the imaginary error function represented here
through the Kummer confluent hyperbolic function. In turn, the second
integral in Eq.~(\ref{eq_44}) is of the following form:
\begin{equation}
\int_{0}^{u}z\,\pFq{1}{1}\!\left(1;\frac{3}{2};az^{2}\right)\mathrm{d}z=
\frac{u^{2}}{2}\pFq{2}{2}\!\left(1,1;\frac{3}{2},2;au^{2}\right).
\label{eq_46}
\end{equation}
Therefore, utilizing Eqs.~(\ref{eq_45}) and (\ref{eq_46}) in Eq.~(\ref{eq_44})
we instantly reconstruct the exact formula for the mean first-passage time
downhill of the harmonic potential as given by Eq.~(\ref{eq_41}).

The second approach is possible due to the application of two integrals. The
first one emerges from Eq.~(\ref{eq_45}) and the fact that $\pFq{0}{0}(;;az^{2})
\!=\!\exp(az^{2})$, which results in
\begin{equation}
\int_{0}^{u}\exp\!\big(az^{2}\big)\,\mathrm{d}z=u\,\pFq{1}{1}\!\left(\frac{1}{2};
\frac{3}{2};au^{2}\right).
\label{eq_47}
\end{equation}
The second integral of the following form
\begin{equation}
\int_{0}^{u}\!z^{\lambda}\exp\!\big(a^{2}z^{2}\big)\,\mathrm{erf}(az)\,\mathrm{d}z
=\frac{2au^{\lambda+2}}{\sqrt{\pi}(\lambda+2)}\,\pFq{2}{2}\!\left(1,\frac{\lambda}
{2}+1;\frac{3}{2},\frac{\lambda}{2}+2;a^{2}u^{2}\right)
\label{eq_48}
\end{equation}
holds for $\mathrm{Re}(\lambda)\!>\!-2$~\cite{Prud1986}. Setting here
$\lambda\!=\!0$ and $a\!=\!\sqrt{b}$, we transform it to the more specific form:
\begin{equation}
\int_{0}^{u}\!\exp\!\big(bz^{2}\big)\,\mathrm{erf}\big(\sqrt{b}z\big)\,\mathrm{d}z
=\sqrt{\frac{b}{\pi}}\,u^{2}\,\pFq{2}{2}\!\left(1,1;\frac{3}{2},2;bu^{2}\right).
\label{eq_49}
\end{equation}
Taking into account the complementary error function $\mathrm{erfc}(z)\!=\!
1-\mathrm{erf}(z)$, we can recast Eq.~(\ref{eq_42}) as follows:
\begin{equation}
\mathcal{T}_{\scriptscriptstyle{\swarrow}}(x_{0}\!\to\!0)=\sqrt{\frac{\pi}
{2D\alpha}}\int_{0}^{x_{0}}\exp\!\left(\frac{\alpha y^{2}}{2D}\right)\mathrm{d}
y-\sqrt{\frac{\pi}{2D\alpha}}\int_{0}^{x_{0}}\exp\!\left(\frac{\alpha y^{2}}
{2D}\right)\mathrm{erf}\!\left(\sqrt{\frac{\alpha}{2D}}y\right)\mathrm{d}y,
\label{eq_50}
\end{equation}
and immediately reproduce, using this formula along with the integrals embodied
by Eqs.~(\ref{eq_47}) and (\ref{eq_49}), the central result displayed in 
Eq.~(\ref{eq_41}).

To demonstrate the correctness of Eq.~(\ref{eq_41}) we will consider the second
technique corresponding to the solution of the partial differential equation
embodied by Eq.~(\ref{eq_13}). We will show that this solution eventually
coincides with Eq.~(\ref{eq_31}) from which the main result for the mean
first-passage time downhill of the harmonic potential $V(x)\!=\!\frac{1}{2}
\alpha x^{2}$ emerges. Thus, in this particular case the backward Fokker-Planck
equation for the survival probability has the following structure:
\begin{equation}
\frac{\partial}{\partial t}Q(t\!\mid\!x)=-\alpha x\frac{\partial}{\partial x}
Q(t\!\mid\!x)+D\frac{\partial^{2}}{\partial x^{2}}Q(t\!\mid\!x).
\label{eq_51}
\end{equation}
As the result of the Laplace transformation made on the above equation, we obtain
that
\begin{equation}
s\,\tilde{Q}(s\!\mid\!x)-Q(0\!\mid\!x_{0})=-\alpha x\frac{\partial}{\partial x}
\tilde{Q}(s\!\mid\!x)+D\frac{\partial^{2}}{\partial x^{2}}\tilde{Q}(s\!\mid\!x).
\label{eq_52}
\end{equation}
Taking into account the initial condition $Q(0\!\mid\!x_{0})\!=\!1$ and defining
the auxiliary function $\tilde{W}(x,s)\!=\!\tilde{Q}(s\!\mid\!x)\!-\!\frac{1}{s}$
with the new variable $z\!=\!\sqrt{\frac{\alpha}{2D}}x$, we can recast
Eq.~(\ref{eq_52}) to the following form:
\begin{equation}
\frac{\partial^{2}}{\partial z^{2}}\tilde{W}(z,s)-2z\frac{\partial}{\partial z}
\tilde{W}(z,s)-\frac{2s}{\alpha}\tilde{W}(z,s)=0.
\label{eq_53}
\end{equation}
The general solution of this equation is known and expressed as the linear
combination of the Hermite function and the Kummer confluent hypergeometric
function. The formal expression of this solution reads
\begin{equation}
\tilde{W}(z,s)=A\,H_{-\frac{s}{\alpha}}(z)+B\,\pFq{1}{1}\!\left(\frac{s}{2\alpha};
\frac{1}{2};z^{2}\right),
\label{eq_54}
\end{equation}
where the numerical parameters $A$ and $B$ have to be determined by imposing the
appropriate boundary conditions. In addition, since the Hermite function is 
related to the Tricomi confluent hypergeometric function, i.e. $H_{\nu}(z)\!=\!
2^{-\nu}\,U\!\left(\frac{\nu}{2},\frac{1}{2},z^{2}\right)$, therefore 
Eq.~(\ref{eq_54}) can be rewritten in the alternate form
\begin{equation}
\tilde{W}(x,s)=\frac{A}{2^{s/\alpha}}\,U\!\left(\frac{s}{2\alpha},\frac{1}{2},
\frac{\alpha x^{2}}{2D}\right)+B\,\pFq{1}{1}\!\left(\frac{s}{2\alpha};
\frac{1}{2};\frac{\alpha x^{2}}{2D}\right),
\label{eq_55}
\end{equation}
in which we have returned to the original variable $x$. In the Laplace domain
the boundary condition $Q(t\!\mid\!\infty)\!=\!1$ becomes $\tilde{Q}(s\!\mid\!
\infty)\!=\!\frac{1}{s}$, which implicates $\tilde{W}(\infty,s)\!=\!0$. From
the two hypergeometric functions in Eq.~(\ref{eq_55}) only the first one
satisfies this property, because in general $\lim_{z\to\infty}U(a,b,z)\!=\!0$, 
whereas $\lim_{z\to\infty}\pFq{1}{1}(a;b;z)\!=\!\infty$ for any $a\!>\!0$. Hence,
\begin{equation}
\tilde{W}(x,s)=\frac{A}{2^{s/\alpha}}\,U\!\left(\frac{s}{2\alpha},\frac{1}{2},
\frac{\alpha x^{2}}{2D}\right),
\label{eq_56}
\end{equation}
where we assume that the parameter $A$ is to be determined by the absorbing
boundary condition $Q(t\!\mid\!x_{\text{a}})\!=\!0$ imposed at the position
$x\!=\!x_{\text{a}}$. As a consequence, we find that $\tilde{W}(x_{\text{a}},s)
\!=\!-\frac{1}{s}$ in the Laplace domain, which allows us to identify the 
constant $A$. Knowing $A$ and demanding that the particle diffusing in the
harmonic potential starts at time $t\!=\!0$ from the initial position $x=x_{0}
\!>\!x_{\text{a}}$, we show that Eq.~(\ref{eq_56}) takes the unambiguous form
\begin{equation}
\tilde{W}(x_{0},s)=-\frac{U\!\left(\frac{s}{2\alpha},\frac{1}{2},\frac{\alpha
x_{0}^{2}}{2D}\right)}{s\,U\!\left(\frac{s}{2\alpha},\frac{1}{2},\frac{\alpha
x_{{\text{a}}}^{2}}{2D}\right)}.
\label{eq_57}
\end{equation}
Armed with this result, we readily obtain the solution of the backward
Fokker-Planck equation for the survival probability in the Laplace domain,
that is
\begin{equation}
\tilde{Q}(s\!\mid\!x_{0})=\frac{1}{s}\left[1-\frac{U\!\left(\frac{s}{2\alpha},
\frac{1}{2},\frac{\alpha x_{0}^{2}}{2D}\right)}{U\!\left(\frac{s}{2\alpha},
\frac{1}{2},\frac{\alpha x_{{\text{a}}}^{2}}{2D}\right)}\right].
\label{eq_58}
\end{equation}
Accordingly, using Eq.~(\ref{eq_10}), we see that the Laplace transform of the
first-passage time distribution is given by
\begin{equation}
\tilde{F}(s\!\mid\!x_{0})=\frac{U\!\left(\frac{s}{2\alpha},
\frac{1}{2},\frac{\alpha x_{0}^{2}}{2D}\right)}{U\!\left(\frac{s}{2\alpha},
\frac{1}{2},\frac{\alpha x_{{\text{a}}}^{2}}{2D}\right)}.
\label{eq_59}
\end{equation}
For the absorbing boundary condition localized in the minimum $x\!=\!0$ of the
harmonic potential, $x_{\text{a}}\!=\!0$. In this peculiar case, the Tricomi
confluent hypergeometric function $U(a,b,0)\!=\!\frac{\Gamma(1-b)}{\Gamma(a-b+1)}$
for $\mathrm{Re}(b)\!<\!1$. Therefore, we show that $U\!\left(\frac{s}{2\alpha},
\frac{1}{2},0\right)\!=\!\sqrt{\pi}\,\Gamma^{-1}\!\left(\frac{s}{2\alpha}+\frac{1}
{2}\right)$, because the Euler gamma function $\Gamma\!\left(\frac{1}{2}\right)
\!=\!\sqrt{\pi}$. In this way, Eq.~(\ref{eq_59}) takes the particular form
\begin{equation}
\tilde{F}(s\!\mid\!x_{0})=\frac{1}{\sqrt{\pi}}\,\Gamma\!\left(\frac{s}{2\alpha}+
\frac{1}{2}\right)U\!\left(\frac{s}{2\alpha},\frac{1}{2},\frac{\alpha\,x^{2}_{0}}
{2D}\right).
\label{eq_60}
\end{equation}
Note, we have already derived this formula in Eq.~(\ref{eq_31}) by the
application of a significantly different method.

Let us now try the third approach consisting of the solution of the
differential equation (\ref{eq_15}) to prove the correctness of the main
formula in Eq.~(\ref{eq_41}) for the mean first-passage time downhill of the
harmonic potential. In the first step, we have to solve the following equation:
\begin{equation}
D\frac{\mathrm{d}^{2}\mathcal{T}(x)}{\mathrm{d}x^{2}}-\alpha x\frac{
\mathrm{d}\mathcal{T}(x)}{\mathrm{d}x}=-1,
\label{eq_61}
\end{equation}
to find its general solution. Lowering the order of the above ordinary
differential equation by substitution $\frac{\partial\mathcal{T}(x)}{\partial x}
\!=\!\mathcal{Y}(x)$ and taking advantage of a standard procedure for solving the 
first-order differential equations (see for example~\cite{Poly2018}), we find
that
\begin{equation}
\mathcal{Y}(x)=A\exp\!\left(\frac{\alpha x^{2}}{2D}\right)-\frac{1}{D}\exp\!\left
(\frac{\alpha x^{2}}{2D}\right)\!\int\exp\!\left(-\frac{\alpha x^{2}}{2D}\right)
\mathrm{d}x,
\label{eq_62}
\end{equation}
where $A$ is the first constant of integration. The second coefficient results
from the subsequent undefined integration, which we will perform in the moment.
For our convenience, let us first determine the undefined integral in
Eq.~(\ref{eq_62}). Its more general form can be found in~\cite{Grad2007},
where 
\begin{equation}
\int\!\mathrm{e}^{-(ax^{2}+2bx+c)}\,\mathrm{d}x=\frac{1}{2}\sqrt{\frac{\pi}{a}}
\exp\!\left(\frac{b^{2}-ac}{a}\right)\mathrm{erf}\!\left(\sqrt{a}x+\frac{b}
{\sqrt{a}}\right)
\label{eq_63}
\end{equation}
for $a\!\ne\!0$. Hence, assuming $b\!=\!c\!=\!0$ and setting $a\!=\!\frac{\alpha}
{2D}$, as well as performing the second integration of Eq.~(\ref{eq_62}), we
obtain
\begin{equation}
\mathcal{T}(x)=A\int\!\exp\!\left(\frac{\alpha x^{2}}{2D}\right)\mathrm{d}x-
\sqrt{\frac{\pi}{2D\alpha}}\int\!\exp\!\left(\frac{\alpha x^{2}}{2D}\right)
\mathrm{erf}\!\left(\sqrt{\frac{\alpha}{2D}}x\right)\mathrm{d}x+B.
\label{eq_64}
\end{equation}
The two integrals appearing in the above expression can be calculated as follows.
The combination of the formula $\int\!\exp\!\left(ax^{2}\right)\mathrm{d}x\!=\!
\frac{1}{2}\sqrt{\frac{\pi}{a}}\mathrm{erfi}(\sqrt{a}x)$ with a representation of
the imaginary error function through the Kummer confluent hypergeometric function,
namely $\mathrm{erfi}(z)\!=\!\frac{2z}{\sqrt{\pi}}\pFq{1}{1}\!\left(\frac{1}{2};
\frac{3}{2};z^{2}\right)$, leads to the first integral in Eq.~(\ref{eq_64}) of
the form
\begin{equation}
\int\exp\!\left(\frac{\alpha x^{2}}{2D}\right)\mathrm{d}x=x\,\pFq{1}{1}\!\left(
\frac{1}{2};\frac{3}{2};\frac{\alpha x^{2}}{2D}\right).
\label{eq_65}
\end{equation}
The second integral in Eq.~(\ref{eq_64}) is a bit more difficult to perform. To
determine it, we use a more general expression, where $\int z^{\alpha-1}
\exp(a^{2}z^{2})\mathrm{erf}(az)\mathrm{d}x\!=\!\frac{2a}{\sqrt{\pi}(\alpha+1)}
z^{\alpha+1}\pFq{2}{2}\!\left(1,\frac{\alpha+1}{2};\frac{3}{2},\frac{\alpha+3}{2};
a^{2}z^{2}\right)$. Here, the result is given by the generalized hypergeomertic
function. Therefore, in the special case of the parameter $\alpha\!=\!1$, we
see that
\begin{equation}
\int\exp\!\left(\frac{\alpha x^{2}}{2D}\right)\mathrm{erf}\bigg(\sqrt{\frac{
\alpha}{2D}}x\bigg)\mathrm{d}x=\sqrt{\frac{\alpha}{2\pi D}}\,x^{2}\,
\pFq{2}{2}\!\left(1,1;\frac{3}{2},2;\frac{\alpha x^{2}}{2D}\right).
\label{eq_66}
\end{equation}
Inserting the last two undefined integrals into Eq.~(\ref{eq_64}) allows us
to recast the general solution of the second-order differential equation
(\ref{eq_61}) in the following form:
\begin{equation}
\mathcal{T}(x)=A\,x\,\pFq{1}{1}\!\left(\frac{1}{2};\frac{3}{2};\frac{\alpha x^{2}}
{2D}\right)-\frac{x^{2}}{2D}\,\pFq{2}{2}\!\left(1,1;\frac{3}{2},2;\frac{
\alpha x^{2}}{2D}\right)+B.
\label{eq_67}
\end{equation}
Hereafter, our task is to determine two unknown integration constants. We will
do that in a few consecutive steps for diffusion proceeding downhill of the
harmonic potential in the present section, whereas the reverse process will be
considered in the subsequent section.

Our primary challenge is to show that the first derivative of the function in 
Eq.~(\ref{eq_67}) with respect to the coordinate $x$ reads
\begin{equation}
\frac{\mathrm{d}\mathcal{T}(x)}{\mathrm{d}x}=\exp\!\left(\frac{\alpha x^{2}}{2D}
\right)\left[A-\sqrt{\frac{\pi}{2D\alpha}}\,\mathrm{erf}\!\left(\sqrt{\frac{\alpha}
{2D}}x\right)\right].
\label{eq_68}
\end{equation}
For this purpose, we utilize the following derivatives of the hypergeometric
functions:
\begin{equation}
\frac{\partial}{\partial z}\pFq{1}{1}(a;b;z)=\frac{a}{b}\pFq{1}{1}(a+1;b+1;z)
\label{eq_69}
\end{equation}
and
\begin{equation}
\frac{\partial}{\partial z}\pFq{2}{2}(a_{1},a_{2};b_{1},b_{2};z)=
\frac{a_{1}a_{2}}{b_{1}b_{2}}\pFq{2}{2}(a_{1}+1,a_{2}+1;b_{1}+1,b_{2}+1;z),
\label{eq_70}
\end{equation}
so that the result obtained from Eq.~(\ref{eq_68}) is as follows:
\begin{align}
\frac{\mathrm{d}\mathcal{T}(x)}{\mathrm{d}x}&=A\!\left[\pFq{1}{1}\!\left(\frac{1}
{2};\frac{3}{2};\frac{\alpha x^{2}}{2D}\right)+\frac{\alpha x^{2}}{3D}\,
\pFq{1}{1}\!\left(\frac{3}{2};\frac{5}{2};\frac{\alpha x^{2}}{2D}\right)\!\right]
\nonumber\\
&-\frac{x}{D}\pFq{2}{2}\!\left(1,1;\frac{3}{2},2;\frac{\alpha x^{2}}{2D}
\right)-\frac{\alpha x^{3}}{6D^{2}}\pFq{2}{2}\!\left(2,2;\frac{5}{2},3;\frac{
\alpha x^{2}}{2D}\right).
\label{eq_71}
\end{align}
The Kummer confluent hypergeometric functions enclosed in the first square
bracket can be expressed by more familiar functions, that is, exponential and
imaginary error functions. To show this, it is enough to employ the integral
representation of these hypergeometric functions:
\begin{equation}
\pFq{1}{1}(a;b;z)=\frac{\Gamma(b)}{\Gamma(a)\,\Gamma(b-a)}\int_{0}^{1}
\mathrm{e}^{zu}\,u^{a-1}(1-u)^{b-a-1}\mathrm{d}u.
\label{eq_72}
\end{equation}
By setting $a\!=\!\frac{1}{2}$, $b\!=\!\frac{3}{2}$, we see that
\begin{equation}
\pFq{1}{1}\!\left(\frac{1}{2};\frac{3}{2};z\right)=\int_{0}^{1}\!
\frac{\mathrm{e}^{zu}}{2\sqrt{u}}\,\mathrm{d}u.
\label{eq_73}
\end{equation}
It appears that the above integral is related to the Dawson function $\mathcal{F}
(x)\!=\!\exp(-x^{2})\int_{0}^{x}\exp(y^{2})\,\mathrm{d}y\!=\!\frac{\sqrt{\pi}}
{2}$\linebreak$\exp(-x^{2})\,\mathrm{erfi}(x)$. In fact, by inserting the new
integration variable $y\!=\!\sqrt{zu}$ and identifying $x\!=\!\sqrt{z}$, we 
find after performing simple calculations that $\int_{0}^{1}u^{-1/2}\exp(zu)\,
\mathrm{d}u\!=\!2z^{-1/2}\exp(z)\mathcal{F}(\sqrt{z})$, from which we next prove
that
\begin{equation}
\pFq{1}{1}\!\left(\frac{1}{2};\frac{3}{2};z\right)=\sqrt{\frac{\pi}{4z}}\,
\mathrm{erfi}(\sqrt{z}),
\label{eq_74}
\end{equation}
where the Dawson function has been replaced by the imaginary error function.  

Using the same integral representation of the Kummer confluent hypergeometric
function in Eq.~(\ref{eq_72}) for $a\!=\!\frac{3}{2}$ and $b\!=\!\frac{5}{2}$,
we obtain that
\begin{align}
\pFq{1}{1}\!\left(\frac{3}{2};\frac{5}{2};z\right)&=\frac{3}{2}\int_{0}^{1}\!
\sqrt{u}\,\mathrm{e}^{zu}\mathrm{d}u=\frac{3}{2}\frac{\mathrm{d}}{\mathrm{d}z}\!
\int_{0}^{1}\!\frac{\mathrm{e}^{zu}}{\sqrt{u}}\,\mathrm{d}u\nonumber\\
&=3\frac{\mathrm{d}}{\mathrm{d}z}\!\left[\sqrt{\frac{\pi}{4z}}\,\mathrm{erfi}
(\sqrt{z})\right],
\label{eq_75}
\end{align}
since the second integral in the first line has the same structure as the
integral in Eq.~(\ref{eq_73}), and has already been determined. Therefore,
knowing that the first derivative $\frac{\mathrm{d}}{\mathrm{d}z}
\mathrm{erfi}(z)\!=\!\frac{2\exp\left(z^{2}\right)}{\sqrt{\pi}}$, we readily
acquire the final result:
\begin{equation}
\pFq{1}{1}\!\left(\frac{3}{2};\frac{5}{2};z\right)=\frac{3\exp(z)}{2z}-
\frac{3\sqrt{\pi}}{4z^{3/2}}\,\mathrm{erfi}(\sqrt{z}).
\label{eq_76}
\end{equation}
Now, let us substitute $z\!=\!\frac{\alpha x^{2}}{2D}$ to Eq.~(\ref{eq_74}) and
Eq.~(\ref{eq_76}), simultaneously multiplying the second equation by
$\frac{\alpha x^{2}}{3D}$. In this way, upon adding these two equations by
sides, we obtain the following relationship:
\begin{equation}
\pFq{1}{1}\!\left(\frac{1}{2};\frac{3}{2};\frac{\alpha x^{2}}{2D}\right)+
\frac{\alpha x^{2}}{3D}\,\pFq{1}{1}\!\left(\frac{3}{2};\frac{5}{2};
\frac{\alpha x^{2}}{2D}\right)=\exp\!\left(\frac{\alpha x^{2}}{2D}\right),
\label{eq_77}
\end{equation}
which is exactly the expression enclosed by the square bracket in
Eq.~(\ref{eq_71}).

A slightly more difficult problem arises with the generalized hypergeometric
functions involved in the second line of this equation. However, it turns out
that in this case we can also find a very helpful relationship between these
two functions to simplify the expression for the derivative of the mean
first-passage time. To this aim, it is enough to utilize the dependence between
contiguous hypergeometric functions which in the original form is
\begin{equation}
bz\pFq{2}{2}(a+1,b+1;c+1,d+1;z)+cd[\pFq{2}{2}(a,b;c,d;z)-
\pFq{2}{2}(a+1,b;c,d;z)]=0,
\label{eq_78}
\end{equation}
and which, after simple manipulation, takes a more useful form for our purposes,
namely:
\begin{equation}
\pFq{2}{2}(a,b;c,d;z)+\frac{b}{cd}\,z\,\pFq{2}{2}(a+1,b+1;c+1,d+1;z)
=\pFq{2}{2}(a+1,b;c,d;z).
\label{eq_79}
\end{equation}
A direct application of the above equation along with the allowed
transformation $\pFq{2}{2}(a+1,b;c,d;z)\!=\!\pFq{2}{2}(b,a+1;c,d;z)$ to the first
generalized hypergeometric function in the second line of Eq.~(\ref{eq_71}) gives
\begin{equation}
\pFq{2}{2}\!\left(1,1;\frac{3}{2},2;\frac{\alpha x^{2}}{2D}\right)
+\frac{\alpha x^{2}}{6D}\pFq{2}{2}\!\left(2,2;\frac{5}{2},3;
\frac{\alpha x^{2}}{2D}\right)=
\pFq{2}{2}\!\left(1,2;\frac{3}{2},2;\frac{\alpha x^{2}}{2D}\right).
\label{eq_80}
\end{equation}
Hereafter, we will try to express the hypergeometric function on the
right-hand side of the above equation by the product of more elementary
functions. To do this, we begin with the integral representation of the
generalized hypergeometric function
\begin{equation}
\pFq{2}{2}(a_{1},a_{2};b_{1},b_{2};z)=\frac{1}{\Gamma(a_{2})}\int_{0}^{\infty}\!
\mathrm{e}^{-u}u^{a_{2}-1}\pFq{1}{2}(a_{1};b_{1},b_{2};zu)\,\mathrm{d}u,
\label{eq_81}
\end{equation}
which holds for $\mathrm{Re}(a_{2})\!>\!0$. In this integral, there appears
the next generalized hypergeometric function, whose integral representation
is given by the following formula:
\begin{equation}
\pFq{1}{2}(a_{1};b_{1},b_{2};z)=\frac{\Gamma(b_{2})}{\Gamma(a_{1})\,
\Gamma(b_{2}-a_{1})}\int_{0}^{1}(1-u)^{b_{2}-a_{1}-1}u^{a_{1}-1}
\pFq{0}{1}(;b_{1};zu)\,\mathrm{d}u,
\label{eq_82}
\end{equation}
provided the condition $\mathrm{Re}(b_{2})\!>\!\mathrm{Re}(a_{1})$ is met.
In turn, the confluent hypergeometric function in the above expression has
the integral representation
\begin{equation}
\pFq{0}{1}(;b_{1};z)=\frac{2\,\Gamma(b_{1})}{\sqrt{\pi}\,\Gamma\!\left(b_{1}-
\frac{1}{2}\right)}\int_{0}^{1}\!\big(1-u^{2}\big)^{b_{1}-3/2}
\cosh(2\sqrt{z}u)\,\mathrm{d}u,
\label{eq_83}
\end{equation}
under the condition that $\mathrm{Re}(b_{1})\!>\!\frac{1}{2}$. Now, we must
solve this hierarchy of integrals to get the intended result, setting
$a_{1}\!=\!1$, $a_{2}\!=\!2$, $b_{1}\!=\!\frac{3}{2}$ and $b_{2}\!=\!2$.
Therefore, we start backward from Eq.~(\ref{eq_83}) and readily obtain that
\begin{equation}
\pFq{0}{1}\!\left(;\frac{3}{2};z\right)=\int_{0}^{1}\!\cosh(2\sqrt{z}u)
\,\mathrm{d}u=\frac{\sinh(2\sqrt{z})}{2\sqrt{z}}.
\label{eq_84}
\end{equation}
Taking into account Eq.~(\ref{eq_82}) and the above outcome, we find that
\begin{equation}
\pFq{1}{2}\!\left(1;\frac{3}{2},2;z\right)=\int_{0}^{1}\!
\pFq{0}{1}\!\left(;\frac{3}{2};zu\right)\mathrm{d}u=
\int_{0}^{1}\frac{\sinh(2\sqrt{zu})}{2\sqrt{zu}}\mathrm{d}u=
\left(\frac{\sinh(\sqrt{z})}{\sqrt{z}}\right)^{\!2}.
\label{eq_85}
\end{equation}
In the last step, we need to return to the first equation (\ref{eq_81}) of the
integral hierarchy and note that
\begin{equation}
\pFq{2}{2}\!\left(1,2;\frac{3}{2},2;z\right)=\int_{0}^{\infty}\!u\,
\mathrm{e}^{-u}\pFq{1}{2}\!\left(1;\frac{3}{2},2;zu\right)\!\mathrm{d}u
=\frac{1}{z}\!\int_{0}^{\infty}\!\mathrm{e}^{-u}\big[\sinh(\sqrt{zu})\big]^{2}
\mathrm{d}u.
\label{eq_86}
\end{equation}
Here, we have encountered the integral which is rather easy to solve. For this
purpose, it is enough to use $\sinh(z)\!=\!\frac{1}{2}(\mathrm{e}^{z}-
\mathrm{e}^{-z})$ and take advantage of the integral $\int_{0}^{z}\exp(
-u^{2})\mathrm{d}u\!=\!\frac{\sqrt{\pi}}{2}\mathrm{erf}(z)$. Then, the result
is such that $\int_{0}^{\infty}\!\mathrm{e}^{-u}[\sinh(\sqrt{zu})]^{2}
\mathrm{d}u\!=\!\frac{1}{2}\sqrt{\pi z}\,\mathrm{e}^{z}\mathrm{erf}(\sqrt{z})$.
Thus, Eq.~(\ref{eq_86}) implies that the final formula for the specific
form of the generalized hypergeometric function on the right-hand side of
Eq.~(\ref{eq_80}) is
\begin{equation}
\pFq{2}{2}\!\left(1,2;\frac{3}{2},2;z\right)=\frac{\sqrt{\pi}\exp(z)\,
\mathrm{erf}(\sqrt{z})}{2\sqrt{z}}.
\label{eq_87}
\end{equation}
This means that after multiplying either side of Eq.~(\ref{eq_80}) by
$\frac{x}{D}$ and inserting into it the above formula, we obtain the following
relationship
\begin{equation}
\frac{x}{D}\,\pFq{2}{2}\!\left(1,1;\frac{3}{2},2;\frac{\alpha x^{2}}{2D}\right)
+\frac{\alpha x^{3}}{6D^{2}}\,\pFq{2}{2}\!\left(2,2;\frac{5}{2},3;\frac{\alpha
x^{2}}{2D}\right)
=\sqrt{\frac{\pi}{2D\alpha}}\exp\!\left(\frac{\alpha x^{2}}
{2D}\right)\mathrm{erf}\!\left(\sqrt{\frac{\alpha}{2D}}x\right).
\label{eq_88}
\end{equation}
This identity and the one given by Eq.~(\ref{eq_77}), when used in
Eq.~(\ref{eq_71}), reproduce the crucial result for the first derivative of the
mean first-passage time embodied by Eq.~(\ref{eq_68}).

\begin{figure}[t]
\centering
\includegraphics[scale=0.3]{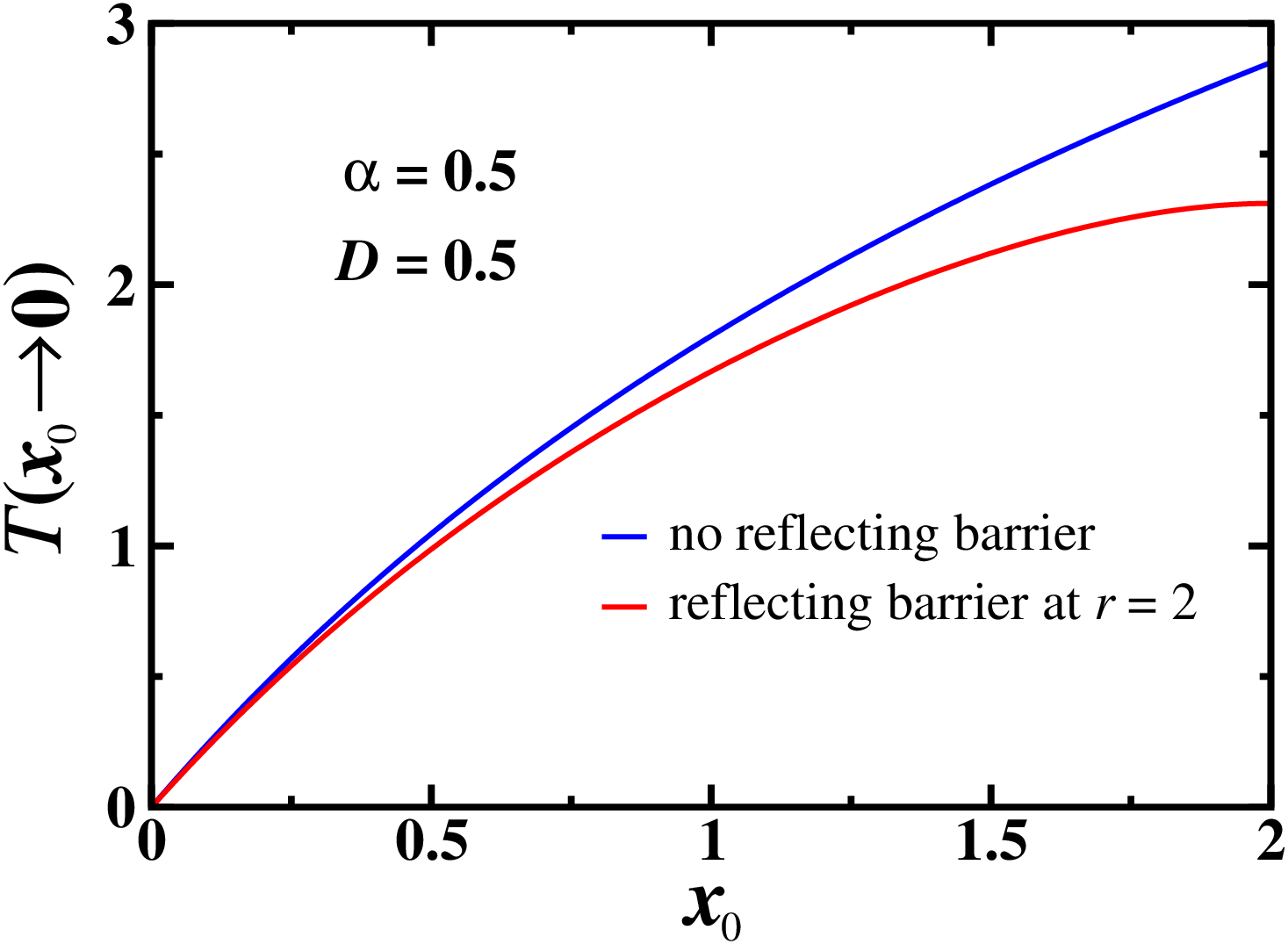}
\caption{The effect of the reflecting barrier on the mean first-passage time
downward of the harmonic potential to the target point at $x\!=\!0$. The blue
line corresponds to the analytic formula in Eq.~(\ref{eq_41}), while the shape
of the red line is described by Eq.~(\ref{eq_90}), where the position of the 
reflecting barrier $r$ coincides with the position $x_{0}\!=\!2$ maximally
distant from the target point. The values of the potential strength $\alpha\!=\!
0.5$ and the diffusion coefficient $D\!=\!0.5$ have been assumed.}
\label{fig3}
\end{figure}

Let us now consider the diffusion downhill of the harmonic potential to the
target point in its minimum $x\!=\!0$. We assume that this point totally absorbs
the particle which initially starts from the position $x_{0}\!>\!0$. In
addition, we define a reflecting point $r\!>\!x_{0}$ to the right of the
initial position (see Fig.~\ref{fig1}). Therefore, the absorbing boundary
condition imposed at $x\!=\!0$ implies that the mean first-passage time
$\mathcal{T}(0)\!=\!0$, if the particle already occupies the minimum $x\!=\!0$
of the harmonic potential. On the other hand, the reflecting boundary condition
makes the first derivative $\frac{\mathrm{d}\mathcal{T}(x)}{\mathrm{d}x}\!=\!0$
at $x\!=\!r$. In this way, we infer from Eq.~(\ref{eq_67}) for $x\!=\!0$ that
the coefficient $B\!=\!0$, since both hypergeometric functions $\pFq{1}{1}
(a;b;0)\!=\!1$ and $\pFq{2}{2}(a_{1},a_{2};b_{1},b_{2};0)\!=\!1$. Taking into
account Eq.~(\ref{eq_68}) and the reflecting boundary condition at $x\!=\!r$,
we readily find that the coefficient
\begin{equation}
A=\sqrt{\frac{\pi}{2D\alpha}}\,\mathrm{erf}\!\left(\sqrt{\frac{\alpha}{2D}}\,r
\right).
\label{eq_89}
\end{equation}
Having determined the integration constants in the solution of the differential
equation (\ref{eq_61}) allows us to show that the mean first-passage time from
$x_{0}\!>\!0$ to the minimum of the harmonic potential at $x\!=\!0$ in the
presence of the reflecting barrier $r\!>\!x_{0}$ is
\begin{equation}
\mathcal{T}_{\scriptscriptstyle{\swarrow}}(x_{0}\!\to\!0)=\sqrt{\frac{\pi x_{0}
^{2}}{2D\alpha}}\,\mathrm{erf}\!\left(\sqrt{\frac{\alpha}{2D}}\,r\right)
\pFq{1}{1}\!\left(\frac{1}{2};\frac{3}{2};\frac{\alpha x_{0}^{2}}{2D}
\right)-\frac{x_{0}^{2}}{2D}\,\pFq{2}{2}\!\left(1,1;\frac{3}{2},2;
\frac{\alpha x_{0}^{2}}{2D}\right).
\label{eq_90}
\end{equation}
In the special case of the reflecting barrier at infinity, i.e. $r\!\to\!\infty$,
the error function $\mathrm{erf}(\infty)\!=\!1$ and hence, we immediately
reproduce the previous result in Eq.~(\ref{eq_41}). Fig.~\ref{fig3} compares
the dependence of the mean first-passage time on the initial position $x_{0}$ of
the particle diffusing downward of the harmonic potential in the absence and
the presence of the reflecting boundary condition. The reflecting barrier $r$
has been assumed to be positioned at the distance $x_{0}\!=\!2$ relative to the
target point shared with the minimum $x\!=\!0$ of the potential. We see that
the reflecting barrier shortens the mean time of diffusion occurring downhill
of the harmonic potential.
\begin{figure}[t]
\centering
\includegraphics[scale=0.3]{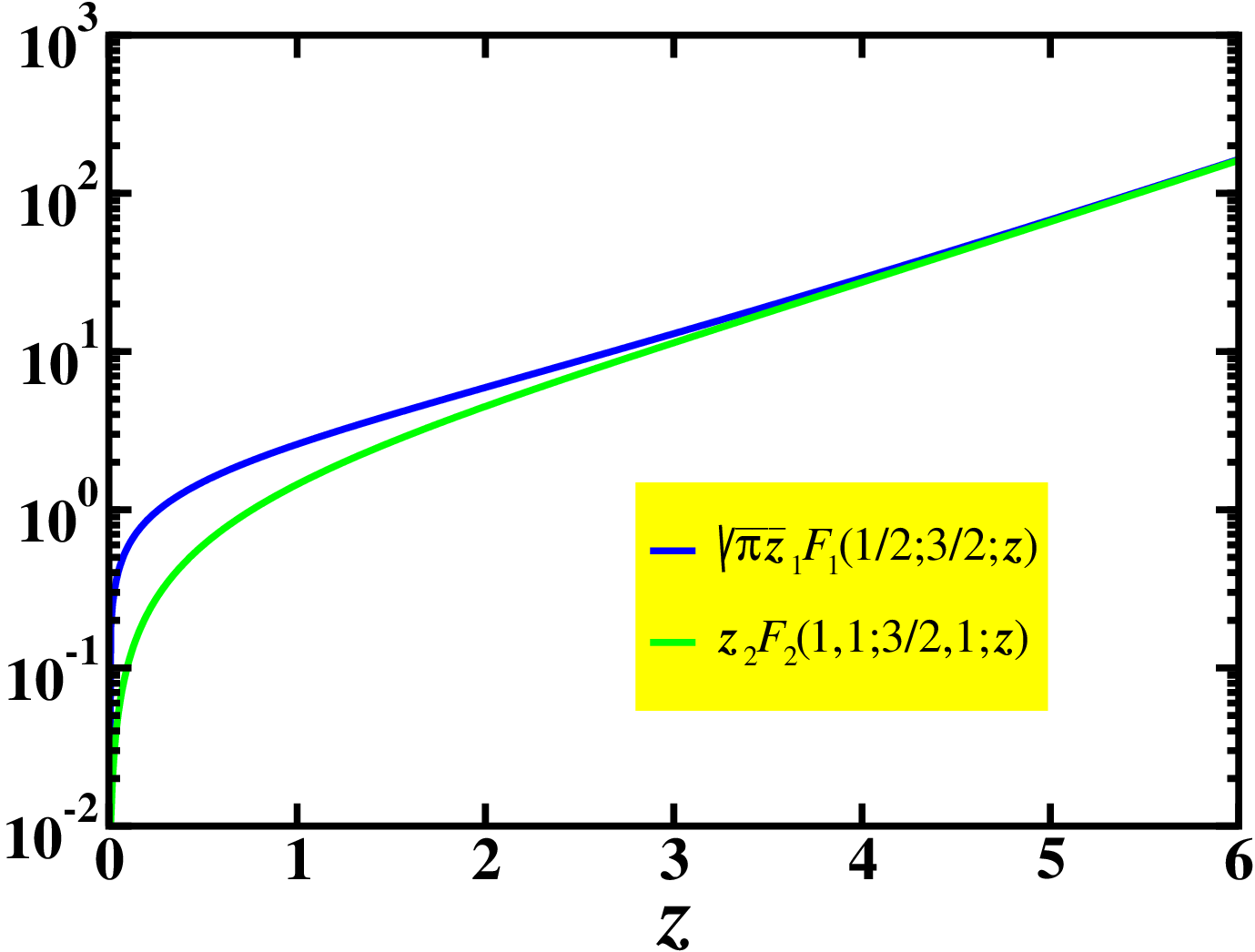}
\caption{Lin-log plot of two components in Eq.~(\ref{eq_41}), the first containing
the Kummer confluent hypergeometric function $\pFq{1}{1}\!\left(\frac{1}{2};\frac{
3}{2};z\right)$ and the second corresponding to the generalized hypergeometric 
function $\pFq{2}{2}\!\left(1,1;\frac{3}{2},2;z\right)$ with $z\!=\!\frac{
\alpha x_{0}^{2}}{2D}$. Both the components rapidly increase to infinity giving
a contribution to the indeterminate form $\infty\!-\!\infty$ if the difference
in the values of these hypergeometric functions is measured.}
\label{fig4}
\end{figure}
\begin{figure}[b]
\centering
\includegraphics[scale=0.25]{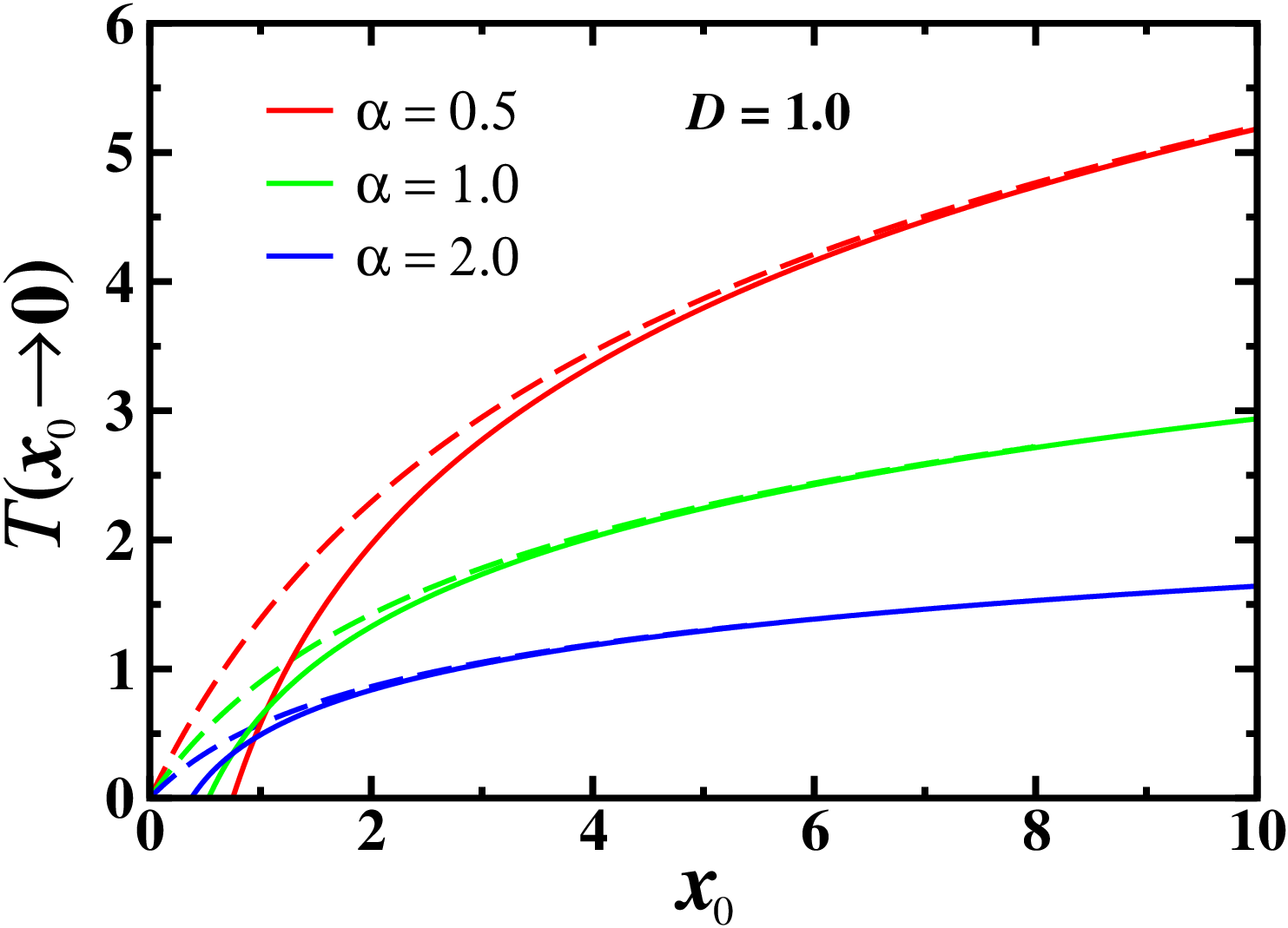}
\caption{Convergence of the asymptotic formula in Eq.~(\ref{eq_93}) for the mean
first-passage time downhill of the harmonic potential (solid lines) to the exact
analytical result (dashed lines) given by Eq.~(\ref{eq_41}). A few values of the
parameter $\alpha$ have been established and the diffusion constant $D\!=\!1.0$
has been assumed.}
\label{fig5}
\end{figure}

The exact expression for the mean first-passage time in Eq.~(\ref{eq_41}) has  
to be complemented with one vital comment that applies to both hypergeometric
functions. These functions rapidly diverge to infinity, even for not so large
values of the distance between the minimum $x\!=\!0$ of the harmonic potential
and the initial position $x_{0}$ of a diffusing particle. This tendency is clearly
depicted in Fig.~\ref{fig4}. For this reason, the mean first-passage time in 
Eq.~(\ref{eq_41}) ceases to be a well-defined quantity for $\frac{\alpha x_{0}
^{2}}{2D}\!\gg\!1$, because it takes an indeterminate form $\infty\!-\!\infty$.
However, we can overcome this difficulty using the asymptotic representation of
the Kummer confluent hypergeometric function.
\begin{equation}
\pFq{1}{1}\!\left(\frac{1}{2};\frac{3}{2};z\right)\propto\frac{\mathrm{e}^{z}}{2z}
-\frac{\mathrm{i}\sqrt{\pi}}{2\sqrt{z}},
\label{eq_91}
\end{equation}
as well as the generalized hypergeometric function,
\begin{equation}
\pFq{2}{2}\!\left(1,1;\frac{3}{2},2;z\right)\propto\frac{\sqrt{\pi}\mathrm{e}^{z}}
{2z^{3/2}}-\frac{\log(4z)+\gamma+\mathrm{i}\pi}{2z}
\label{eq_92}
\end{equation}
for $\lvert z\rvert\!\to\!\infty$, where $\mathrm{i}\!=\!\sqrt{-1}$ and
$\gamma\!\approx\!0.5772$ are the well-known imaginary unit and the aforementioned
Euler-Mascheroni constant, respectively. Therefore, the asymptotic representation
of Eq.~(\ref{eq_41}) takes the following form:
\begin{equation}
\mathcal{T}_{\scriptscriptstyle{\swarrow}}(x_{0}\!\to\!0)\propto\frac{1}{\alpha}
\log\!\left(\sqrt{\frac{2\alpha}{D}}x_{0}\right)+\frac{\gamma}{2\alpha}.
\label{eq_93}
\end{equation}
Figure~\ref{fig5} shows how the asymptotic formula in Eq.~(\ref{eq_93}) for the
mean first-passage time downhill of the harmonic potential converges to the
exact result given by Eq.~(\ref{eq_41}) for long distances from the target
in the minimum of the potential. In this case, we have selected three different
values of the stiffness parameter $\alpha$.

On the other hand, in the vicinity of the target point $x\!=\!0$, that is, when
$\frac{\alpha x_{0}^{2}}{2D}\!\ll\!1$, we can approximate Eq.~(\ref{eq_41}) due to
the Taylor expansion of both hypergeometric functions. Indeed, if $\lvert z\rvert
\!\to\!0$, then
\begin{equation}
\pFq{1}{1}(a;b;z)\propto1+\frac{az}{b},\;\;\mathrm{while}\;\;\pFq{2}{2}(a_{1},a_{2};
b_{1},b_{2};z)\propto1+\frac{a_{1}a_{2}}{b_{1}b_{2}}z,
\label{eq_94}
\end{equation}
and hence
\begin{equation}
\mathcal{T}_{\scriptscriptstyle{\swarrow}}(x_{0}\!\to\!0)\simeq\sqrt{\frac{\pi}
{2D\alpha}}x_{0}.
\label{eq_95}
\end{equation}
In this way, we have retrieved Eq.~(\ref{eq_30}) displaying the linear dependence
of the mean first-passage time on the initial position $x_{0}\!>\!0$ in the close
proximity of the target point anchored in the minimum of the harmonic potential.
This result also suggests that Eq.~(\ref{eq_29}) should be equivalent to 
Eq.~(\ref{eq_41}). Indeed, inserting the integral representation of the error
function, 
\begin{equation}
\mathrm{erf}(az)=\frac{2az}{\sqrt{\pi}}\int_{0}^{1}\mathrm{e}^{-a^{2}z^{2}u^{2}}
\mathrm{d}u
\label{eq_96}
\end{equation}
with $a\!=\!\sqrt{\frac{\alpha}{2D}}\lvert x_{0}\rvert$ and $z\!=\!
\left(\mathrm{e}^{2\alpha t}-1\right)^{-1/2}$ into Eq.~(\ref{eq_29}), and
changing the order of integration, we can directly convert this equation into
Eq.~(\ref{eq_41}) via the utilization of Eq.~(\ref{eq_19}).
\begin{figure}[b]
\centering
\includegraphics[scale=0.25]{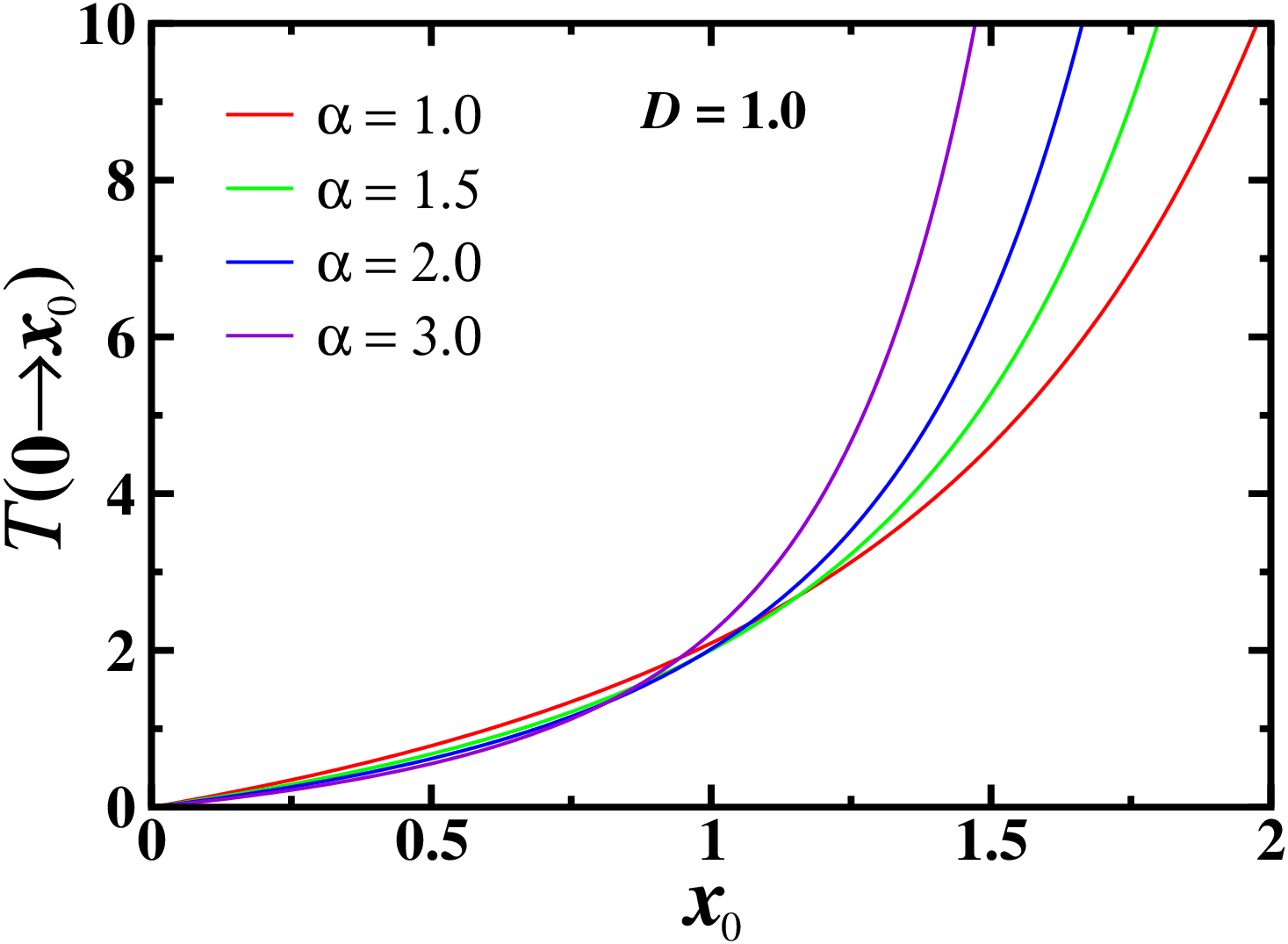}
\caption{Mean first-passage time for diffusion uphill of the harmonic potential
from its minimum at $x\!=\!0$ to the target point at $x_{0}\!>\!0$. A few
values of the parameter $\alpha$ have been established and the diffusion
coefficient $D\!=\!1.0$ has been assumed.}
\label{fig6}
\end{figure}

\section*{4.~Mean first-passage time for diffusion upward of harmonic potential}

In this section we concentrate our attention on the diffusion uphill of the
harmonic potential $V(x)\!=\!\frac{1}{2}\alpha x^{2}$ from its minimum at
$x\!=\!0$ to the target localized at the position $x_{0}\!>\!0$. Out of the
methods collected in Sec.~2, we will apply those manifested in Eqs.~(\ref{eq_15})
and (\ref{eq_17}). Such a strategy based on the two independent methods allows
us to verify the correctness of the final result.

We begin from the second equation which for the harmonic potential reads:
\begin{equation}
\mathcal{T}_{\scriptscriptstyle{\nearrow}}(0\!\to\!x_{0})=\frac{1}{D}
\int_{0}^{x_{0}}\!\mathrm{d}y\,\exp\!\left[\frac{\alpha y^{2}}{2D}\right]\!
\int_{-\infty}^{y}\!\exp\!\left[-\frac{\alpha z^{2}}{2D}\right]\mathrm{d}z.
\label{eq_97}
\end{equation}
To take advantage of the already used relationships, let us transform the last
integral in the above expression as follows:
\begin{equation}
\int_{-\infty}^{y}\exp\!\left[-\frac{\alpha z^{2}}{2D}\right]\mathrm{d}z=
\int_{-\infty}^{\infty}\exp\!\left[-\frac{\alpha z^{2}}{2D}\right]\mathrm{d}z-
\int_{y}^{\infty}\exp\!\left[-\frac{\alpha z^{2}}{2D}\right]\mathrm{d}z.
\label{eq_98}
\end{equation}
Here, the first integral on the right-hand side corresponds to the Gaussian 
integral $\int_{-\infty}^{\infty}\exp\!\big(\!-\frac{\alpha z^{2}}
{2D}\big)\mathrm{d}z\!=\!\sqrt{\frac{2\pi D}{\alpha}}$, while the second
integral gives the complementary error function $\mathrm{erfc}(z)\!=\!
1-\mathrm{erf}(z)$, that is $\int_{y}^{\infty}\exp\!\big(\!-\frac{\alpha z^{2}}
{2D}\big)\mathrm{d}z\!=\!\sqrt{\frac{\pi D}{2\alpha}}\,\mathrm{erfc}\!
\left(\sqrt{\frac{\alpha}{2D}}y\right)$. Inserting these two results into
Eq.~(\ref{eq_98}), we obtain that
\begin{equation}
\int_{-\infty}^{y}\exp\!\left[-\frac{\alpha z^{2}}{2D}\right]\mathrm{d}z=
\sqrt{\frac{\pi D}{2\alpha}}\left[1+\mathrm{erf}\!\left(\sqrt{\frac{\alpha}{2D}}y
\right)\right].
\label{eq_99}
\end{equation}
On the other hand, substituting this integral into Eq.~(\ref{eq_97}) leads to
the following partial result:
\begin{equation}
\mathcal{T}_{\scriptscriptstyle{\nearrow}}(0\!\to\!x_{0})=\sqrt{\frac{\pi}{2
\alpha D}}\int_{0}^{x_{0}}\!\exp\!\left[\frac{\alpha y^{2}}{2D}\right]\!
\mathrm{d}y+\sqrt{\frac{\pi}{2\alpha D}}\int_{0}^{x_{0}}\!\exp\!
\left[\frac{\alpha y^{2}}{2D}\right]\mathrm{erf}\!\left(\sqrt{\frac{\alpha}{2D}}
y\right)\mathrm{d}y.
\label{eq_100}
\end{equation}
Both of the above integrals have already appeared in Eqs.~(\ref{eq_47}) and
(\ref{eq_49}) of the previous section. By using them in Eq.~(\ref{eq_100}), we
readily find the exact formula for the mean first-passage time uphill of the
harmonic potential from its minimum in $x\!=\!0$ to the target point at
$x_{0}\!>\!0$, namely
\begin{equation}
\mathcal{T}_{\scriptscriptstyle{\nearrow}}(0\!\to\!x_{0})=
\sqrt{\frac{\pi x_{0}^{2}}{2D\alpha}}\,\pFq{1}{1}\!\left(\frac{1}{2};\frac{3}{2};
\frac{\alpha x_{0}^{2}}{2D}\right)+\frac{x_{0}^{2}}{2D}\,\pFq{2}{2}\!\left(1,1;
\frac{3}{2},2;\frac{\alpha x_{0}^{2}}{2D}\right).
\label{eq_101}
\end{equation}
The dependence of the mean first-passage time on the distance from the initial
position at $x\!=\!0$ to the target point at $x\!=\!x_{0}$ for a particle
diffusing upward of the harmonic potential is shown in Fig.~\ref{fig6}.
Exemplary values of the strength $\alpha$ of the harmonic potential have been
chosen and the diffusion coefficient $D\!=\!1.0$ has been assumed.
\begin{figure}[bh]
\centering
\includegraphics[scale=0.3]{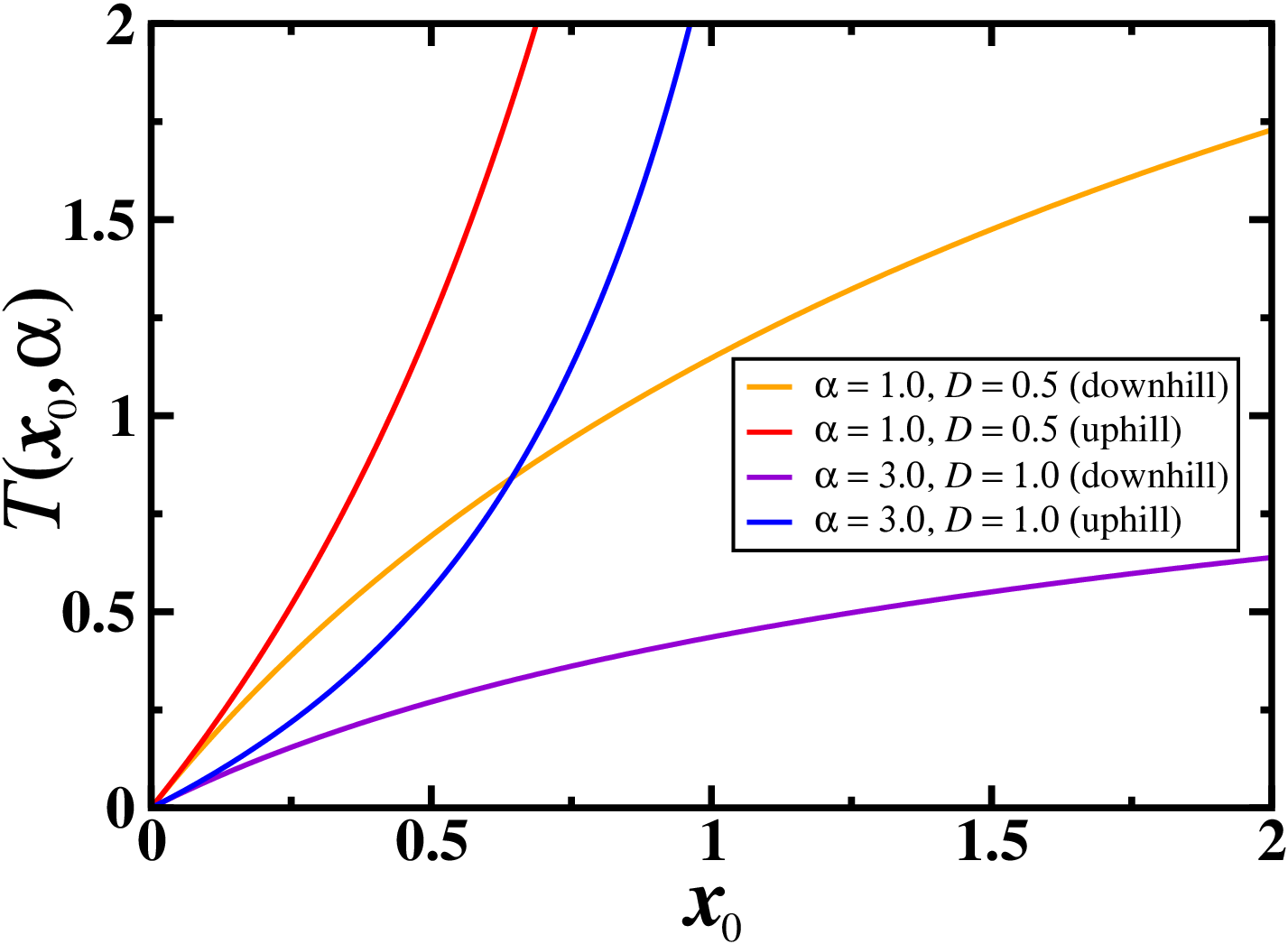}
\caption{Comparison of the mean first-passage times upward and downward of the
harmonic potential. In the first case a particle diffuses from the minimum of
the harmonic potential at $x\!=\!0$ to the target point localized at
$x_{0}\!>\!0$. In the second case it starts at $x_{0}\!>\!0$ and diffuses to
the point in the minimum $x\!=\!0$ of the harmonic potential. Two different
values of the parameter $\alpha$ have been selected and two distinct values of
the diffusion coefficient have been assumed.}
\label{fig7}
\end{figure}

The different method that confirms the result in Eq.~(\ref{eq_101}) relates to
the second-order differential equation (\ref{eq_61}). We again posit that the
absorbing boundary condition is imposed on the target at $x_{0}\!>\!0$.
In addition, the reflecting barrier $r\!>\!0$ is somewhere between the
minimum $x\!=\!0$ of the harmonic potential and this target point,
therefore $0\!<\!r\!<\!x_{0}$. We also assume for the moment that the
initial position of the particle is not precisely determined but must
be included in the range between $r$ and $x_{0}$. The two boundary
conditions imply $\mathcal{T}(x_{0})\!=\!0$ and
$\frac{\mathrm{d}\mathcal{T}(x)}{\mathrm{d}x}\!\Big|_{x=r}\!=\!0$,
respectively, for the mean first-passage time of the particle that
initially occupies the target point and its first derivative over
the coordinate $x$ that disappears at the reflecting point.

Combining these two boundary conditions with Eqs.~(\ref{eq_67}) and
(\ref{eq_68}), we readily show that the coefficient
\begin{equation}
A=\sqrt{\frac{\pi}{2D\alpha}}\,\mathrm{erf}\!\left(\sqrt{\frac{\alpha}{2D}}
\,r\right),
\label{eq_102}
\end{equation}
while the second coefficient
\begin{equation}
B=\frac{x_{0}^{2}}{2D}\,\pFq{2}{2}\!\left(1,1;\frac{3}{2},2;\frac{\alpha
x_{0}^{2}}{2D}\right)-\sqrt{\frac{\pi x_{0}^{2}}{2D\alpha}}\,\mathrm{erf}\!
\left(\sqrt{\frac{\alpha}{2D}}\,r\right)\pFq{1}{1}\!\left(\frac{1}{2};
\frac{3}{2};\frac{\alpha x_{0}^{2}}{2D}\right).
\label{eq_103}
\end{equation}
The last step of our calculations is to make the substitution of the above 
constants into the solution of the second-order differential equation
(\ref{eq_61}), which is embodied by Eq.~(\ref{eq_67}) of the previous section.
In this way we finally obtain that
\begin{align}
\mathcal{T}(x)&=\sqrt{\frac{\pi}{2D\alpha}}\,\mathrm{erf}\!
\left(\sqrt{\frac{\alpha}{2D}}\,r\right)\!\left[x\,\pFq{1}{1}\!\left(\frac{1}{2};
\frac{3}{2};\frac{\alpha x^{2}}{2D}\right)-x_{0}\,\pFq{1}{1}\!\left(\frac{1}{2};
\frac{3}{2};\frac{\alpha x_{0}^{2}}{2D}\right)\!\right]\nonumber\\
&+\frac{1}{2D}\!\left[x_{0}^{2}\,\pFq{2}{2}\!\left(1,1;\frac{3}{2},2;
\frac{\alpha x_{0}^{2}}{2D}\right)-x^{2}\,\pFq{2}{2}\!\left(1,1;\frac{3}{2},2;
\frac{\alpha x^{2}}{2D}\right)\right].
\label{eq_104}
\end{align}
Now we can take advantage of three properties corresponding to the error function
for which $\mathrm{erf}(-z)\!=\!-\mathrm{erf}(z)$, hence $\mathrm{erf}(-\infty)
\!=\!-1$ since $\mathrm{erf}(\infty)\!=\!1$, and the hypergeometric functions
for which $\pFq{1}{1}(a;b;0)\!=\!1$ and $\pFq{2}{2}(a_{1},a_{2};b_{1},b_{2};0)
\!=\!1$. Applying these rules to Eq.~(\ref{eq_104}) and assuming that the
reflecting barrier is pushed back to minus infinity, while the particle
initiates its diffusive motion from the minimum $x\!=\!0$ of the harmonic
potential, we directly reproduce the main result exposed in Eq.~(\ref{eq_101}).
However, if the reflecting barrier overlaps with the initial position of the
diffusing particle at $x\!=\!0$, then $\mathrm{erf}(0)\!=\!0$ and consequently
Eq.~(\ref{eq_104}) simplifies to the more compact form:
\begin{equation}
\mathcal{T}_{\scriptscriptstyle{\nearrow}}(0\!\to\!x_{0})=
\frac{x_{0}^{2}}{2D}\,\pFq{2}{2}\!\left(1,1;\frac{3}{2},2;\frac{\alpha x_{0}^{2}}
{2D}\right).
\label{eq_105}
\end{equation}

However, comparing the main result included in Eq.~(\ref{eq_41}) with that given
by (\ref{eq_101}), we notice the only difference in the middle sign, while the
rest structure of these formulas is exactly the same. Accordingly, the mean
first-passage time uphill of the harmonic potential has to be longer than the
one in the opposite direction. This is consistent with the fact that a diffusive
motion is slower uphill, while faster downhill of the confining harmonic
potential. Figure~\ref{fig7} collects results for the mean first-passage times
downward and upward of the harmonic potentials assuming various values of the
stiffness parameter $\alpha$ and the diffusion constant $D$.

\section*{5.~Conclusions}

The motivation for writing this paper was born out of the lack of complete
analytical expressions for the mean first-passage time downward and upward of the
harmonic potential. We have obtained these exact results using a few disparate
methods, which allowed us to validate their correctness. Figure~(\ref{fig7})
contains a collection of four exemplary graphs plotted in line with the formulas
embodied by Eqs.~(\ref{eq_41}) and (\ref{eq_101}) for two different values of the
parameter $\alpha$ and the diffusion coefficient $D$. In the case of diffusion
downhill of the harmonic potential, the mean first-passage time is shortened due
to the increase of the potential strength and additionally the value of the
diffusion constant. Surprisingly, when diffusive dynamics take place upward of
the harmonic potential then we can observe a substantially different tendency.
For small distances between the initial position of a diffusing particle at the
minimum $x\!=\!0$ of the potential and the target point at $x_{0}\!>\!0$, the
mean first-passage time turns out to be shorter for higher values of the parameter
$\alpha$ than lower ones. This is possible as long as the diffusion coefficient
is relatively larger with respect to the larger $\alpha$ than the smaller one.
After exceeding a certain distance that separates the starting and target points,
we again observe a characteristic elongation of the mean first-passage time with
an increase of the parameter $\alpha$ (see Fig.~\ref{fig6}). We connect this
effect with the course of confluent and generalized hypergeometric functions
depending on the change in the location of a target point relative to the
starting point (see Fig.~\ref{fig4}). Recall that the combination of these two
functions defines a full expression for the mean first-passage time uphill of
the harmonic potential.

We have also shown that the mean first-passage time downward of the harmonic
potential depends on the difference of confluent and generalized
hypergeometric functions. This, in turn, raises the problem of determining its
value for longer distances between the initial position at $x_{0}\!>\!0$ and
the target that coincides with the minimum $x\!=\!0$ of the harmonic potential.
We have also argued that to overcome this problem, it is necessary to use the
asymptotic expansion, see Eq.~(\ref{eq_93}), of the exact formula for the mean
first-passage time in Eq.~(\ref{eq_41}). The validity of such an approach is
confirmed in Fig.~\ref{fig5}, where a good convergence of both functions at
relatively larger distances is detected.

The new rigorous result obtained in this paper concerns the presence of
the reflecting boundary condition in the harmonic potential and its influence on
the mean first-passage time of a particle diffusing both downhill, as well as 
uphill of this potential. For example, we have shown in Fig.~\ref{fig3} that
the confinement of a particle between the absorbing target and the reflecting
barrier shortens its mean time to hit this target during diffusion downward
of the harmonic potential.

In this paper, we have explored the Ornstein-Uhlenbeck process in order to
obtain the sort of new exact results for the mean first-passage time problem.
They complement the collection of other rigorous solutions that have been 
obtained for physical processes occurring in the harmonic potential, such as
classical and quantum harmonic oscillators. We hope that our paper will provide
inspiration for further studies aimed at finding exact results for similar 
physical problems.

\end{document}